\documentclass[review, 10pt]{article}         

\usepackage{mathtools}      
\usepackage[utf8]{inputenc} 
\usepackage[T1]{fontenc}    
\usepackage[hidelinks]{hyperref}       
\usepackage{url}            
\usepackage{microtype}      
\usepackage{amsfonts}       
\usepackage{nicefrac}       
\usepackage{amsthm}         
\usepackage{amssymb}        
\usepackage{bbm}            
\usepackage{booktabs}       
\usepackage{ctable}         
\usepackage{caption}        
\usepackage{multirow}       
\usepackage{tabularx,booktabs}  
\usepackage{graphicx}       
\usepackage{subcaption}
\usepackage{float}          
\usepackage{cleveref}       
\usepackage{cite}
\usepackage{array}          
\usepackage{dirtytalk}      
\usepackage{algorithm} 
\usepackage{algpseudocode} 
\usepackage{amssymb} 
\usepackage{cleveref}
\usepackage{xcolor}
\usepackage{graphicx}
\usepackage{subcaption}
\usepackage{listings}
\usepackage{color}
\usepackage{array}
\usepackage[left=2.75cm,right=2.75cm, bottom=3.00cm, top=3.00cm]{geometry}
\usepackage{soul}
\usepackage[parfill]{parskip}
\usepackage{lineno}
\usepackage{setspace}
\usepackage{rotating}
\usepackage[makeroom]{cancel}
\usepackage{array}

\newcolumntype{P}[1]{>{\centering\arraybackslash}p{#1}}

\allowdisplaybreaks

\newcommand\tab[1][0.5cm]{\hspace*{#1}}         
\newcolumntype{Y}{>{\centering\arraybackslash}X} 

\providecommand{\keywords}[1]{
  \textbf{\textit{Keywords---}} #1
}

\theoremstyle{definition} 

\title{\bf 
Mesh d-refinement: a data-based computational framework to account for complex material response
}
\author{
Sacha Wattel$^{1}$, 
Jean-François Molinari$^{1}$,
Michael Ortiz$^{2,3}$,
Joaquin Garcia-Suarez$^{1}$\footnote{Corresponding author: \texttt{joaquin.garciasuarez@epfl.ch}} 
}
\date{ \small
$^{1}$Institute of Civil Engineering, Institute of Materials, \'{E}cole Polytechnique F\'{e}d\'{e}rale de Lausanne (EPFL), CH 1015 Lausanne, Switzerland\\
$^{2}$Division of Engineering and Applied Science, California Institute of Technology, 1200 E. California Blvd., Pasadena, CA 91125, USA\\
$^{3}$Hausdorff Center for Mathematics, Universität Bonn, Endenicher Allee 60, 53115 Bonn, Germany}
\begin{document}

\maketitle

\begin{abstract}
Model-free data-driven computational mechanics (DDCM) is a new paradigm for simulations in solid mechanics. The modeling step associated to the definition of a material constitutive law is circumvented through the introduction of an abstract phase space in which, following a pre-defined rule, physically-admissible states are matched to observed material response data (coming from either experiments or lower-scale simulations).
%
In terms of computational resources, the search procedure that performs these matches is the most onerous step in the algorithm.
%
%
%
One of the main advantages of DDCM is the fact that it avoids regression-based, bias-prone constitutive modeling.
However, many materials do display a simple linear response in the small-strain regime while also presenting complex behavior after a certain deformation threshold.
Motivated by this fact,
we present a novel refinement technique that turns regular elements (equipped with a linear-elastic constitutive law) into data-driven ones if they are expected to surpass the threshold known to trigger material non-linear response. %
We term this technique ``data refinement'', ``d-refinement'' for short. %
Starting from an initially regular FEM mesh, the proposed algorithm detects where the refinement is needed and iterates until all elements presumed to display non-linearity become data-driven ones. Insertion criteria are discussed. The scheme is well-suited for simulations that feature non-linear response in relatively small portions of the domain while the rest remains linear-elastic. The method is validated against a traditional incremental solver (i.e., Newton-Raphson method) and we show that the d-refinement framework can outperform it in terms of speed at no loss of accuracy.
We provide an application that showcases the advantage of the new method: bridging scales in architected metamaterials. 
\end{abstract}


\section{Introduction}
\label{Sec:Introduction}

Data-driven computational mechanics (DDCM) is a rapidly-expanding research field. Data-driven solvers were originally devised to deal with problems in small-strains statics \cite{Trent_1,Conti_Mueller_Ortiz}; extensions to dynamics \cite{Trent_2,Amith,DDNSR}, large deformations \cite{Auriane_1,Conti_Mueller_Ortiz_2,DD_foam} and dissipative (inelastic) material response \cite{Eggersmann,Pietro,Pietro_2,Kostas_2,Kostas_1} followed. 

The data-driven paradigm relies on using ``observed information'' directly, avoiding the fitting/calibration exercise that turns these observations into a fully-defined mathematical function (``material constitutive model''). Thus, part of the efforts in the field have been focused on generating those datasets either from experimental observations \cite{LEYGUE} or from microstructural simulations \cite{Kostas_2,Kostas_1,DD_foam}. Further efforts are currently underway to maximize its numerical efficiency \cite{EGGERSMANN_2,tensor_voting,Kostas_1,Korzeniowski}. The mathematical foundations of the method have been solidly laid out both in small strains \cite{Conti_Mueller_Ortiz} and large deformations \cite{Conti_Mueller_Ortiz_2}.


Arguably, the main benefit of the DDCM paradigm is that it enables circumventing the definition of intricate phenomenological laws, which are required to describe complex material response during numerical simulations \cite{multi_scale_ML}. However, most materials do display a simple, consistent linear-elastic behavior as long as they remain in the small-strain regime, after which either non-linear or inelastic phenomena are expected to kick in. Here, we consider that ``complex response'' is anything other than linear and elastic.
As linear-elastic simulations are both straightforward to implement and computationally efficient, it is therefore desirable to stick to them to solve as much of the simulation domain as possible. For this, one needs to devise a framework that can detect when an element surpasses a deformation threshold, 
and, once that trigger is indeed detected, 
proceed to enrich that element with a data-driven formulation. This would also redound on simulation speed, as it would reduce the number of phase-space searches, which remains the most time-consuming step in the algorithm, despite the fact that they are parallelizable.

A recent article of Yang et al. \cite{Method_1} has been the first, to our knowledge, that combined data-driven and traditional elements in the same mesh. They used their coupling technique to analyze structures in which part of the main domain remains linear-elastic while a portion of it weakens due to an external factor (corrosion). 
Taking advantage of this coupling scheme, 
we introduce a new kind of mesh refinement technique that aims to improve both numerical efficiency and accuracy while maintaining the advantages of DDCM: ``data refinement'', ``d-refinement'' for short, that automatically turns regular elements into data-driven ones where and when necessary. 
This eases the computational implementation, as it avoids intricacies associated with numerically resolving explicitly either material non-linearities, inelastic mechanisms or both. 
D-refinement also improves accuracy: where constitutive modeling is more arduous, it replaces assumption-ridden phenomenological descriptions by data, while keeping the model in the region where it performs best (infinitesimal strain), thus also partially avoiding errors associated with dealing with a discrete material dataset \cite{Trent_1}. See \Cref{Sec:D_ref_more} for further discussion on this point. 
%


We will show that the d-refinement technique does not require load stepping in the absence of inelasticity. Since it combines both FEM and DD, the definition of the constants that metrize the space in DDCM \cite{Trent_1,Kostas_1} is straightforward in d-refinement: we can make them equal to the elastic constants of the FEM linear material.

The d-refinement framework forges a better union between model-driven and data-driven ways of computing. It aspires to compete with incremental solvers based on tangent operators as the tool-of-choice when it comes to perform mechanics calculations that include either non-linear behavior or inelasticity or both. Furthermore, if the dataset comes from lower-scale simulations, d-refinement can also be regarded as an optimized $FE^2$ method \cite{FEYEL}, in which (a) the microstructural response is obtained off-line before the simulation \cite{Kostas_1,DD_foam} and (b) microstructure RVE response is directly used at the element level only when its linear-elastic range is exhausted. 

The paper is structured as follows: \Cref{Sec:methods} reviews the basics of DDCM and the ``static'' FEM-DD implementation, and introduces the d-refinement solving procedure. \Cref{Sec:validation} validates the methodology via comparisons to conventional incremental solvers, we compare two kinds of systems: 3D trusses and 2D plane-stress elements. \Cref{Sec:application} presents an application of d-refinement to multiscale analysis. Starting from a material dataset, we use efficient data-driven simulations to characterize the mechanical response of the unit cell of an architected metamaterial, in terms of both linear-elastic constants and a dataset that includes non-linear response. Then, this information is used to study the K-field in a cracked macroscale sample \cite{toughness_metameterials}.  

\section{Methods} \label{Sec:methods}

\subsection{Data-driven computational mechanics}


DDCM poses the solid mechanics boundary-value problem in an abstract phase space of work-conjugate variables (e.g., strain and stress), in which each element state is defined by a point in a ``local'' phase space, and the overall system phase space is the Cartesian product of that of its elements.
The field equations (equilibrium and compatibility), along with the boundary conditions, represent a manifold embedded in phase space. The infinite set that contains all possible points that satisfy these constraints is termed $\mathrm{E}$, and referred to as the ``physically-admissible set''. Traditionally, the constitutive law would amount to a graph in this space \cite{Conti_Mueller_Ortiz}, classical solutions of the boundary-value problem would correspond to the intersection of the aforementioned manifold and the graph. However, in DDCM, the material behavior is represented by a ``material discrete set'', termed $\mathrm{D}$. As the intersection of admissible set $\mathrm{E}$ and $\mathrm{D}$ is unlikely, this concept is replaced by the notion of ``transversality'' \cite{Conti_Mueller_Ortiz} between $\mathrm{E}$  and $\mathrm{D}$.

The FEM-DD coupling scheme that we are to use and to introduce in the next section converges to the original DD implementation \cite{Trent_1} when the whole mesh becomes data-driven; hence, for the sake of brevity, we introduce the coupling scheme algorithm directly in next section.  


\subsection{FEM-DD coupling}

\subsubsection{Introduction}

We begin with a brief digression concerning terminology.
Model-driven is a fair description of the current paradigm, which relies on phenomenological models of material response to close the problem: boundary conditions, force equilibrium and displacement compatibility equations need to be supplemented with a force-deformation relation. 
Nevertheless, using the acronym ``MD'' to refer to model-driven is deemed confusing, as both in the mechanics and in the material science communities it is associated with ``molecular dynamics'', a field where data-driven approaches are also possible \cite{DD_MD}. 
%
%
We would rather use, at least in the context of solid and structural mechanics, the name ``DD-FEM coupling'' in lieu of ``DD-MD coupling''.
Small caveats aside, the term is self-explanatory and, we believe, will be readily understood by researchers in any of the fields mentioned before. 
In a recent contribution by Korzeniowski and Weinberg \cite{DD_foam}, the authors refer to the conventional DDCM way of computing as ``DD-FEM'' as it is ``data-driven in a finite-element mesh''. However, since we tend to associate the finite-element method not only with the discretization of the domain, but also with a way to model materials (that is why we would speak of ``non-linear finite elements''), we believe our terminology to be appropriate. 

The coupling implementation \cite{Method_1} divides the set of all elements in the numerical model ($S$) into two subsets: $S_2$ includes the indices of DD elements, while $S_1$ contains those of the regular ones ($S_2 \cap S_1 = \emptyset$ and $S_2 \cup S_1 = S$). The size of each subset, $| \cdot|$, is defined as the number of elements in it, thus $|S_2| + |S_1| = N_e$, where $N_e$ is the total number of elements in the mesh. 
The concept of phase space distances remain unchanged for DD elements: 
first, a point in the local phase space $Z_e$ (of the $e$-th element, $e \in S_2$) is 
$\boldsymbol{z}_e = \{ \boldsymbol{\epsilon}_e, \, \boldsymbol{\sigma}_e \} \in \mathbb{R}^{2N_c}$, where $N_c$ is the number of relevant components of stress/strain (one for uni-axial elements, three for plane problems, six for 3D problems, and nine for 3D problems in generalized continua \cite{Kostas_2}). This local phase space is equipped with a metric defining a norm for each point as

\begin{align}
    |\boldsymbol{z}_{e}|^2
    =
    \frac{1}{2} \, \mathbb{C}_e  \boldsymbol{\epsilon}_e \cdot \boldsymbol{\epsilon}_e
    +
    \frac{1}{2} \, \mathbb{C}_e^{-1}  \boldsymbol{\sigma}_e \cdot \boldsymbol{\sigma}_e \, ,
\end{align}

where $\mathbb{C}_e$ is a symmetric positive-definite matrix of constants not necessarily related to any material property.  

Given that the global phase space is $Z = Z_1 \times Z_2 \ldots \times Z_{|S_2|}$, the norm associated to $\boldsymbol{z} \in Z$ can be taken as

\begin{align}
    |\boldsymbol{z}|
    =
    \left(
    \sum_{e \in S_2}
    {w_e}
    |\boldsymbol{z}_{e}|
    \right)^{1/2}
    \, ,
\end{align}

where $w_e$ is the volume of the $e$-th element. The latter in turn provides a notion of distance between two points, $\boldsymbol{z}$ and $\boldsymbol{y}$, in phase space:

\begin{align} \label{eq:distance}
    d(\boldsymbol{z}, \, \boldsymbol{y})
    =
    |\boldsymbol{z} - \boldsymbol{y}| \, .
\end{align}

Note that the solution of traditional finite-element analysis also represents a point in a phase space, and hence the global solution of the system that contains both kinds of elements can be represented in a larger phase space (i.e., $Z_1 \times Z_1 \times \ldots \times Z_e$).

By means of the constitutive law, assumed linear and elastic, we can write $\boldsymbol{\sigma}_e = \mathbb{D}_e \boldsymbol{\epsilon}_e$ for every element s.t. $e \in S_1$, with $\mathbb{D}_e \in \mathbb{R}^{N_c \times N_c}$ containing the regular elastic constants of the $e$-th element. 
Using the discretized compatibility relation, one can express the strain of an element in terms of the global nodal displacement vector, $\boldsymbol{u} \in \mathbb{R}^{N_{\text{dof}}}$ ($N_{\text{dof}}$ is the total number of degrees of freedom in the mesh),  

\begin{align}
    \boldsymbol{\epsilon}_e 
    =
    \boldsymbol{B}_e \boldsymbol{u} 
    \qquad
    \text{for }
    e=1,\ldots, N_e
    \, ,
\end{align}

$\boldsymbol{B}_e \in \mathbb{R}^{N_c \times N_{\text{dof}}}$ encapsulates the nodal connectivity and the shape functions used to perform nodal interpolation within each element. The global equilibrium equation is expressed as

\begin{align}
    \sum_{e \in S_2} 
    w_e
    \boldsymbol{B}_e^{\top}
    \boldsymbol{\sigma}_e 
    +
    \sum_{e \in S_1} 
    w_e
    \boldsymbol{B}_e^{\top}
    \mathbb{D}
    \boldsymbol{B}_e \boldsymbol{u}
    =
    \boldsymbol{f} \, ,
\end{align}
where $\boldsymbol{f} \in $ $\mathbb{R}^{N_{\text{dof}}}$ is the nodal external force vector. 

Notice that the displacements of the regular elements enter the nodal equilibrium equations along with the stress of the DD ones. 
Using Lagrange multipliers, $\boldsymbol{\eta} \in \mathbb{R}^{N_{\text{dof}}}$, these equilibrium constraints must be enforced concurrently with the distance minimization between points in the admissible set, $\boldsymbol{z} \in \mathrm{E}$, and the material set, $\boldsymbol{z}^* \in \mathrm{D} = \mathrm{D}_1 \times \mathrm{D}_2 \times \ldots \times \mathrm{D}_{|S_2|}$, for the DD portion of the mesh \cite{Trent_1}. Thus, we introduce a functional 

\begin{align}
    \Pi 
    =
    d^2(\boldsymbol{z},\, \boldsymbol{z}^*) 
    + 
    \boldsymbol{\eta} 
    \cdot
    \left(
    \sum_{e \in S_2} 
    \boldsymbol{B}_e^{\top}
    \boldsymbol{\sigma}_e 
    +
    \sum_{e \in S_1} 
    \boldsymbol{B}_e^{\top}
    \mathbb{D}
    \boldsymbol{B}_e \boldsymbol{u}
    -
    \boldsymbol{f}
    \right) \, .
\end{align}


Using compatibility and enforcing the stationarity of the said functional renders a problem that can be solved iteratively, until the state of the system that minimizes the distance between the solution and the admissible points in the material dataset is attained. The corresponding Euler-Lagrange equations yield, after minor term re-arrangement,

\begin{subequations} \label{eq:stationarity}
\begin{align}
    \sum_{e \in S_1} 
    w_e
    \boldsymbol{B}_e^{\top}
    \mathbb{D}_e
    \boldsymbol{B}_e \boldsymbol{u}
    -
    \sum_{e \in S_2} 
    w_e
    \boldsymbol{B}_e^{\top}
    \mathbb{C}_e
    \boldsymbol{B}_e \boldsymbol{\eta}
    &=
    \sum_{e \in S_2} 
    w_e
    \boldsymbol{B}_e^{\top}
    \mathbb{C}_e
    \boldsymbol{\epsilon}^*_e \, , \label{eq:stationarity_u} 
    \\
    \sum_{e \in S_1} 
    w_e
    \boldsymbol{B}_e^{\top}
    \mathbb{D}_e
    \boldsymbol{B}_e \boldsymbol{u}
    +
    \sum_{e \in S_2} 
    w_e
    \boldsymbol{B}_e^{\top}
    \mathbb{C}_e
    \boldsymbol{B}_e \boldsymbol{\eta}
    &=
    \boldsymbol{f} 
    -
    \sum_{e \in S_2} 
    w_e
    \boldsymbol{B}_e^{\top}
    \boldsymbol{\sigma}^*_e
    \label{eq:stationarity_f}
    \, ,
\end{align}
\end{subequations}

where we use the superindex ``$*$'' to highlight elements in the material dataset $\mathrm{D}$. Given $ \boldsymbol{z}^*_e = \left( \boldsymbol{\epsilon}^*_e , \, \boldsymbol{\sigma}^*_e \right) \in \mathrm{D}_e$ $\forall e \in S_2$, \cref{eq:stationarity} can be solved for $\boldsymbol{u}$ and $\boldsymbol{\eta}$, and then physically-admissible strain and stresses for the DD elements can be computed as \cite{Method_1}

\begin{subequations} \label{eq:projection_E}
\begin{align} 
    \boldsymbol{\sigma}_e 
    &=
    \boldsymbol{\sigma}_e^*
    +
    \mathbb{D}_e \boldsymbol{B}_e \boldsymbol{\eta} \, , \\
    \boldsymbol{\epsilon}_e 
    &=
    \boldsymbol{B}_e \boldsymbol{u} \, .     
\end{align}
\end{subequations}

The subsequent solving of \cref{eq:stationarity} and \cref{eq:projection_E} is likened to a projection operation that takes the phase-space point $\boldsymbol{z}^*$ in $\mathrm{D}$ and projects it onto the admissible set:  $P_E \boldsymbol{z}^* \in \mathrm{E}$. 

The next step of the algorithm involves projecting back onto the material set, what is achieved through an element-wise search that picks the point in $\mathrm{D}$ closest to the previously obtained $ \boldsymbol{z} \in \mathrm{E}$, and it is symbolically expressed as $P_D \boldsymbol{z} \in \mathrm{D}$. The necessary notion of distance between points is provided by \cref{eq:distance}. In essence, this defines the simplest data-driven solver, which allows fixed-point iterations from the $k$-th admissible DD solution $\boldsymbol{z}^{k}$ to the next one, i.e.,  $\boldsymbol{z}^{(k+1)} = P_E P_D \boldsymbol{z}^{(k)}$. 
The algorithm yields the final solution at the $n$-th iteration if all the DD elements are assigned the same set of data points twice in a row, i.e., $\boldsymbol{z}^{*(n)} = P_D P_E \boldsymbol{z}^{*(n-1)} = \boldsymbol{z}^{*(n-1)}$.

\textcolor{black}{The coupling is expected to converge with respect to the dataset size since the linear-elastic finite elements can be thought as DD ones in which $\boldsymbol{z} = \boldsymbol{z}^* $ always, and thus the theorems developed for pure DD meshes \cite{Conti_Mueller_Ortiz,Conti_Mueller_Ortiz_2} apply.}

\subsubsection{Implementation}

\begin{algorithm}[H]
\caption{Fixed-point algorithm for DD-FEM coupling}\label{alg:coupling}
\begin{algorithmic}

\State \textbf{Require:} $\forall e = 1 , \ldots, N_e$, compatibility matrices $\boldsymbol{B}_e$; $\forall e \in S_1$, $\mathbb{D}_e$ matrices containing elastic constants; $\forall e \in S_2$, local datasets $\mathrm{D}_e$ and $\mathbb{C}_e$ matrices containing distance constants; $\forall i = 1 , \ldots, N_{\text{dof}}$, external forces $f_i$ or boundary conditions. 

\State (i) Set $k=0$. Initial data assignation:
\For{$  e \in S_2$ }\\
\tab Choose $\boldsymbol{z}_e^{*(0)} =  (\boldsymbol{\sigma}_e^{*(0)}, \boldsymbol{\epsilon}_e^{*(0)}) \in \mathrm{D}_e$ randomly
\EndFor

\State (ii) Project onto $\mathrm{E}$: \\
Solve in \cref{eq:stationarity} for $\boldsymbol{u}^{(k)}$ and $\boldsymbol{\eta}^{(k)}$
\For{$ e \in S_2$ }\\
\tab Compute $\boldsymbol{\epsilon}_e$ and  $\boldsymbol{\sigma}_e$ from \cref{eq:projection_E}
\EndFor

\State (iii) Project onto $\mathrm{D}$: 
\For{$ e \in S_2$ }\\
\tab Choose $\boldsymbol{z}_e^{*(k+1)} \in \mathrm{D}_e$ closest to $\boldsymbol{z}_e^{(k)} \in \mathrm{E}_e$
\EndFor

\State (iv) Convergence test:
\If{ $\boldsymbol{z}_e^{(k+1)} = \boldsymbol{z}_e^{(k)}$ $ \forall e \in S_2$} \\
\tab Final displacement field $\boldsymbol{u} = \boldsymbol{u}^{(k)}$ \textbf{exit}
\Else \\
\tab $k \leftarrow k+1$, \textbf{goto} (ii)
\EndIf

\end{algorithmic}
\end{algorithm}

\Cref{alg:coupling} presents the pseudo-code implementation of a coupled FEM-DD solver.

\subsection{D-refinement}

\subsubsection{Motivation} \label{Sec:D_ref_more}

As discussed in the introduction, we foresee data-driven simulations in which some elements represent material behavior that does display a meaningful, well-defined, linear-elastic range. 
Performing phase-space searches for elements that do not abandon the small-strain regime appears unnecessary, as no extra accuracy would be gained at the cost of potential errors associated to dataset finite size. Since DDCM requires no intersection between the sets but just transversality, there is always some potential error that is introduced every time that a material datum is assigned to an element \cite{Trent_1}. In summary, the phase space searches should be limited to regions that clearly depart from linearity, what will be a small portion of the total if the problem displays localization like, e.g., stark stress concentrations or shear banding. 
\textcolor{black}{Once again, let us state that ``complex material response'' in the context of this text refers to anything beyond linear elasticity.}

Let us also remark that the searches become more involved and time-consuming as the dimensionality of the space increases. From 1D requiring only one component of stress and strain to define the per-element 2D phase space, to 3D elements requiring six of each in a 12D phase space, the resources invested in performing the projection onto $\mathrm{D}$ scale rapidly. Of course, the size of $\mathrm{D}$ itself also plays a major role, but this issue can be offset via smart searching procedures \cite{Trent_1,EGGERSMANN_2}, e.g., employing the k-d tree algorithm.

\subsubsection{Implementation details}
\label{Sec:implementation_details}

For maximum efficiency, the raw material dataset should be sifted to get rid of points that represent linear response. For the upcoming simulations, we defined limit stresses $\sigma_{\text{lim}}$ in terms of some combination of the stress components (e.g., mean stress) and chose to include each datum $\boldsymbol{z}_e$ whenever $f(\boldsymbol{z}_e) > 0.8 \, \sigma_{\text{lim}}$, where $f$ represents the aforementioned stress combination. The reason for using a relaxed inclusion is leaving enough wiggle room in the dataset to account also for almost-non-linear response.

Likewise, to take into account the uncertainty associated with the definition of the threshold itself, we choose to switch to DD whenever $f(\boldsymbol{z}) > 0.9 \, \sigma_{\text{lim}}$.

\begin{figure}
    \centering
    \includegraphics[width=0.75\textwidth]{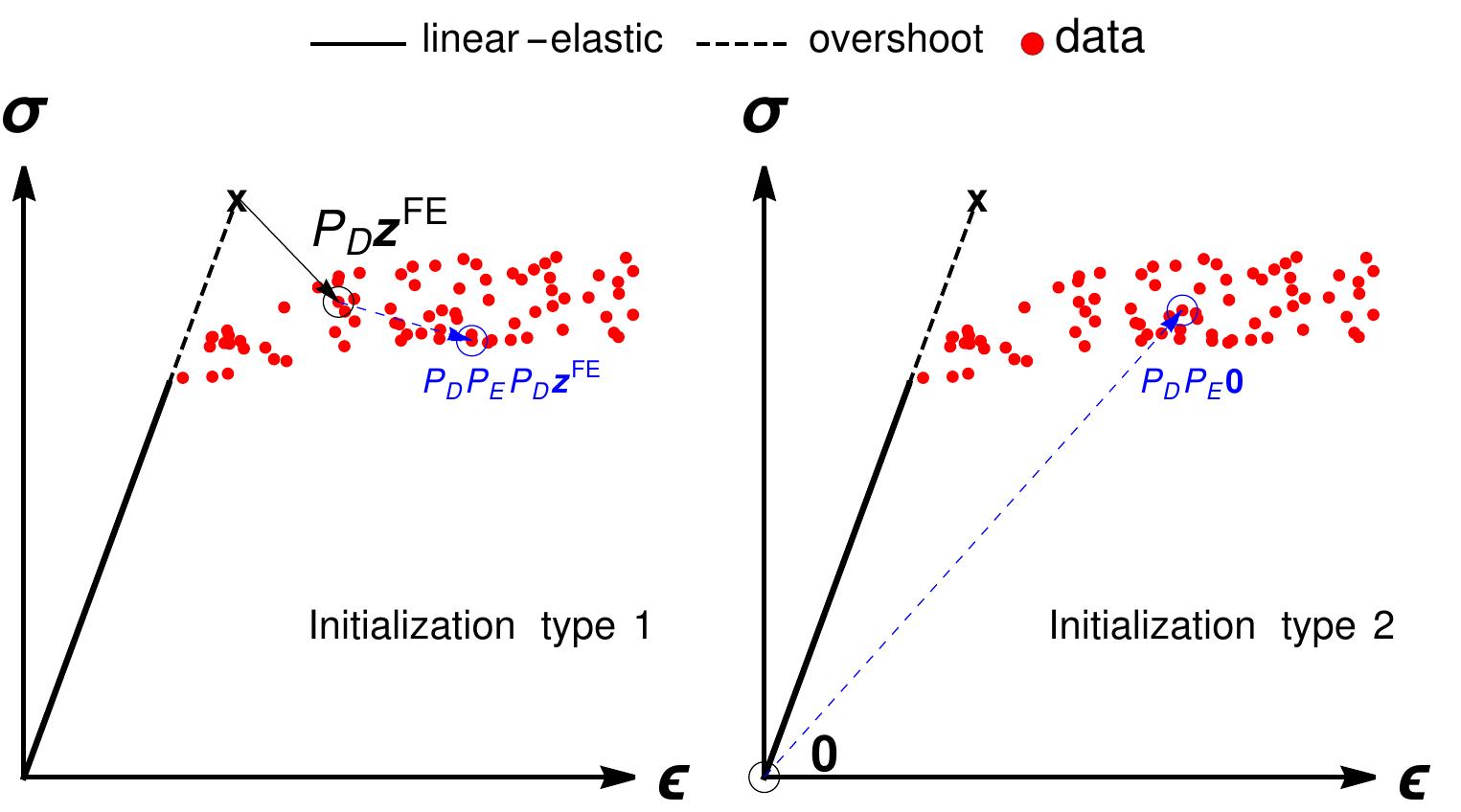}
    \caption{Schematic representation of DD element initialization possibilities on a 2D phase space. The first circle (black) represents the initialization point and the second one (blue) the point in the dataset that the algorithm converges to in the next iteration (see that each method may pick a different datum after the iteration). }
    \label{fig:initilization}
\end{figure}

When a regular element is swapped by a DD one, the strain-stress state of the latter can be initialized in different ways (\Cref{fig:initilization}). 
\begin{itemize}
    \item Method \# 1: The last state of the regular element, call it $\boldsymbol{z}^{FE}_e$, can be swapped by its closest point in $\mathrm{D}$, i.e., by $\boldsymbol{z}_e^* = \text{argmin}~d(\boldsymbol{z}_e^{FE} , \, \mathrm{D})$, or, using projection notation $\boldsymbol{z}_e^* = P_D \boldsymbol{z}_e^{FE}$. 
    This approach initializes the element at a proper location in phase space, at the expense of performing an extra search. 
    \item Method \# 2: Simply initialize the new DD element state at the origin, i.e., $\boldsymbol{z}_e^* = 0 \in \mathbb{R}^{2N_c}$. This method has the advantage of avoiding the initial projection, but could cause the element to follow a phase-space trajectory different from the one in method \#2 in subsequent steps. This approach appears more adequate for history-dependent materials, but these are out of the scope of this text. Herein, we will show that this initialization procedure is faster (since it avoids extra phase-space searches) but can lead to excessive refinement when used without adequate incremental loading.  
\end{itemize}

Incremental loading is unavoidable in some circumstances as, e.g., large deformation analysis (to recompute geometrical stiffness changes associated to shape evolution), in the presence of dynamics (time marching), or inelasticity (history-dependent material response). Nevertheless, when dealing with quasi-static elastic yet non-linear response, the load could be applied at once and the phase-space searches would take care of finding the most convenient element states. We will show that this is indeed the case, but that the final solution can depend on the choice of aforementioned initialization methods.

\begin{algorithm}
\caption{D-refinement framework } \label{alg:d-refinement}
\begin{algorithmic}

\State \textbf{Require:} $\forall e = 1 , \ldots, N_e$, compatibility matrices $\boldsymbol{B}_e$, $\mathbb{D}_e$ matrices containing elastic constants, local datasets $\mathrm{D}_e$ and $\mathbb{C}_e$ matrices containing distance constants; $\forall i = 1 , \ldots, N_{\text{dof}}$, final external forces $F_i$ or boundary conditions. 

\State (i) Set $j=1$, $S_2^{(1)} = \emptyset$ and $S_1^{(1)} = \{ 1, \, 2, \, 3, \ldots , N_e \}$.

\State (ii) Incremental loading. Initialize $l=1$
\For{$l \le N_{\text{steps}}$} \\
Set loading level $\boldsymbol{f} = l/N_{\text{steps}} \boldsymbol{F}$
\If{$|S_2^{(j)}| = 0$}\\ 
\tab \tab Set $\boldsymbol{\eta} = 0$ and solve for $\boldsymbol{u}$ in \cref{eq:stationarity_f}. 
\Else
\tab \State (ii.a) Initialize $k = 0$
\State (ii.b) Project onto $\mathrm{E}$: \\
\tab \tab Solve in \cref{eq:stationarity} for $\boldsymbol{u}^{(k)}$ and $\boldsymbol{\eta}^{(k)}$
\For{$  e \in S_2$ }\\
\tab \tab \tab Compute $\boldsymbol{\epsilon}_e$ and  $\boldsymbol{\sigma}_e$ from \cref{eq:projection_E}
\EndFor

\State (ii.c) Project onto $\mathrm{D}$: 
\For{$  e \in S_2^{(j)}$ }\\
\tab \tab \tab Choose $\boldsymbol{z}_e^{*(k+1)} \in \mathrm{D}_e$ closest to $\boldsymbol{z}_e^{(k)} \in \mathrm{E}_e$
\EndFor

\State (ii.d) Convergence test:
\If{ $\boldsymbol{z}_e^{(k+1)} = \boldsymbol{z}_e^{(k)}$ $\forall e \in S_2^{(j)}$} \\
\tab \tab \tab Final displacement field $\boldsymbol{u} = \boldsymbol{u}^{(k)}$ \textbf{goto} (iii)
\Else \\
\tab \tab \tab $k \leftarrow k+1$, \textbf{goto} (ii.b)
\EndIf
\EndIf

\State (iii) Check for linear elements above threshold:\\ 
\tab Set $S_1^{(j+1)} = S_1^{(j)}$, $S_2^{(j+1)} = S_2^{(j)}$
\For{ $ e \in S_1^{(j)}$ }\\
\tab Compute element strain and/or stress: 
$\boldsymbol{\epsilon}_e$ and $\boldsymbol{\sigma}_e$ 
\tab \If{threshold criterion}\\
\tab \tab \tab Append $e$ to $S_2^{(j+1)}$, drop it from $S_1^{(j+1)}$\\
\tab \tab \tab Initial assignation:
          either
          choose $\boldsymbol{z}_e^{*(k)} \in \mathrm{D}_e$ closest to $(\boldsymbol{\epsilon}_e, \, \boldsymbol{\sigma}_e)$
          or pre-defined $\boldsymbol{z}_e^{*} \in \mathrm{D}_e$
\tab \EndIf
\EndFor

\State (vi) Global convergence condition
\If{$|S_2^{(j)}| = |S_2^{(j-1)}|$}\\
\tab \tab All DD elements converged (or no need for DD), no need of further refinement   \textbf{exit}
\EndIf

$l \leftarrow l+1$

\EndFor

\end{algorithmic}
\end{algorithm}

\section{Verification} \label{Sec:validation}

Two verification exercises are presented in this section. For both, an underlying constitutive law - from which a synthetic database is generated - is assumed to exist, as this allows reaching solutions with a traditional iterative solver, here specifically Newton-Raphson (NR). This iterative solution is then considered as exact in order to estimate the accuracy of the data-driven or d-refined solutions. The two cases are used to study different behaviors of the method. In the first one, constant-stress triangular (CST) elements are used under the plane stress assumption. The main topics of interest are the wall time speed-up, the spread of the refinement and the influence of the initiation of newly refined elements. The second one is a 3D truss-beam discretized with axial bar elements. This is used to study the accuracy and convergence of the method with respect to the number of data points, the influence of noise in the dataset and the impact of the load stepping procedure.

\subsection{2D plane-stress elements: plate with hole}
\label{Sec:validation_plate}

We study first the capabilities of the one-load-step d-refinement procedure. We create a dataset using the per-element phase-space trajectories of the incremental NR solver, and then 
(a) we compare to NR solver in terms of simulation speed assuming no incremental loading in the d-refinement case. The effect of changing the number of elements in the dataset is considered. 
(b) We also verify the accuracy of d-refinement via comparison to NR, and 
(c) compare the two initialization methods (no incremental loading in the d-refinement case) in terms of accuracy (comparison to NR) and computational time. For each initialization method, the proportion of elements that become DD and how many of them reach the same state as in the NR solution is considered. 
Lastly, (d) we assess the impact of widely distributed non-linearity (i.e., higher load, and hence less linear elements and more refinement).

%
The phase-space searches are carried out in parallel using six processors as long as the total number of DD elements is above twelve. This problem is implemented in a Mathematica notebook \cite{Mathematica}, which can be downloaded from on-line repositories (see supplementary material section).

\subsubsection{System}

\begin{figure}
    \centering
    \begin{subfigure}[a]{.58\linewidth}
    \includegraphics[width=\linewidth]{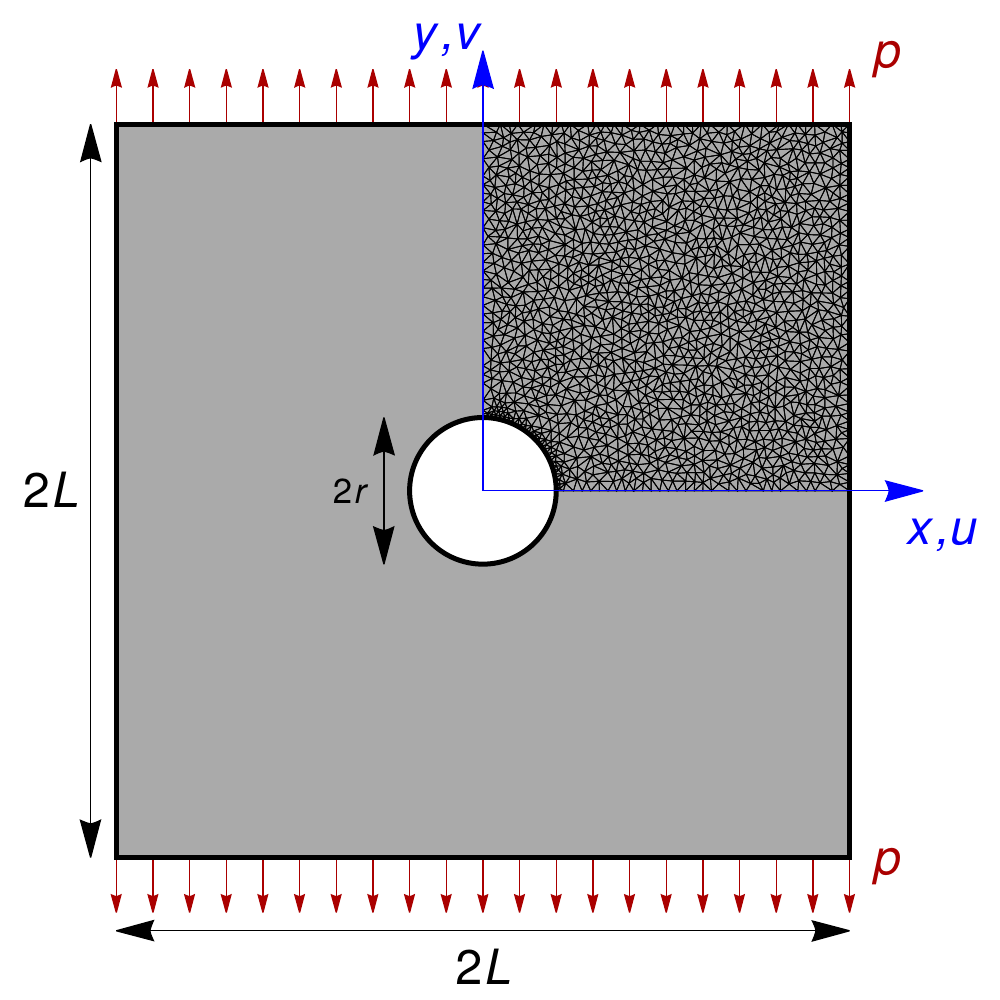}
    \caption{\Large (a)}
    \label{fig:plate_mesh_results_a}
    \end{subfigure}
    \begin{subfigure}[a]{.38\linewidth}
    \includegraphics[width=\linewidth]{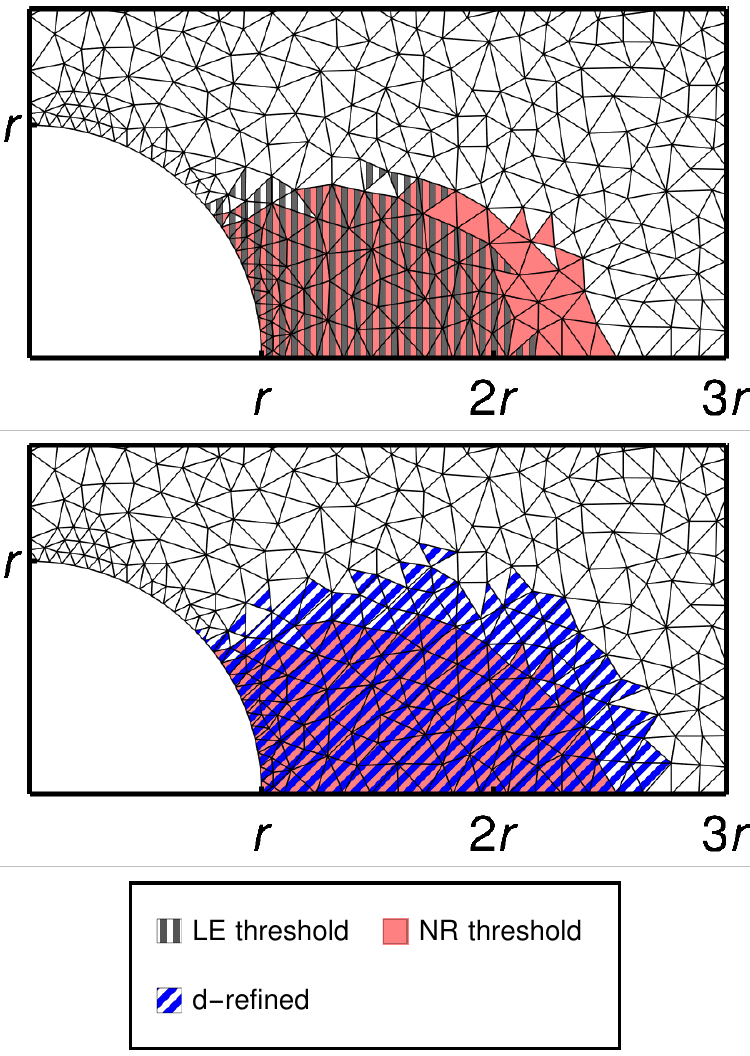}
    \caption{\Large (b)}
    \label{fig:plate_mesh_results_b}
    \end{subfigure}
    \caption{System and simulation's main outcome (results corresponding to $\text{tol} = 10^{-3}$ and |D| = 886 elements). (a) Scheme of the system, including loading and reduced mesh (taking advantage of symmetry). (b) Results: elements above $\sigma_{\text{lim}}$ in linear-elastic simulations (vertically-hatched gray), those above the same threshold in Newton-Raphson simulations (solid pink) and elements that switched to DD in the course of the simulation (diagonally-hatched blue).}
    \label{fig:plate_mesh_results}
\end{figure}

We resort to a classic 2D benchmark exercise: a thin square plate (side length $ L = 0.2 \, \mathrm{m}$) with a circular hole (radius $r = 0.02 \, \mathrm{m}$) loaded in tension on two opposite edges, see \Cref{fig:plate_mesh_results}(a). The applied traction $p$ equals $ 100 \, \mathrm{MPa}$.

\begin{figure}
    \centering
    \begin{subfigure}[a]{.45\linewidth}
    \centering
    \includegraphics[width=\linewidth]{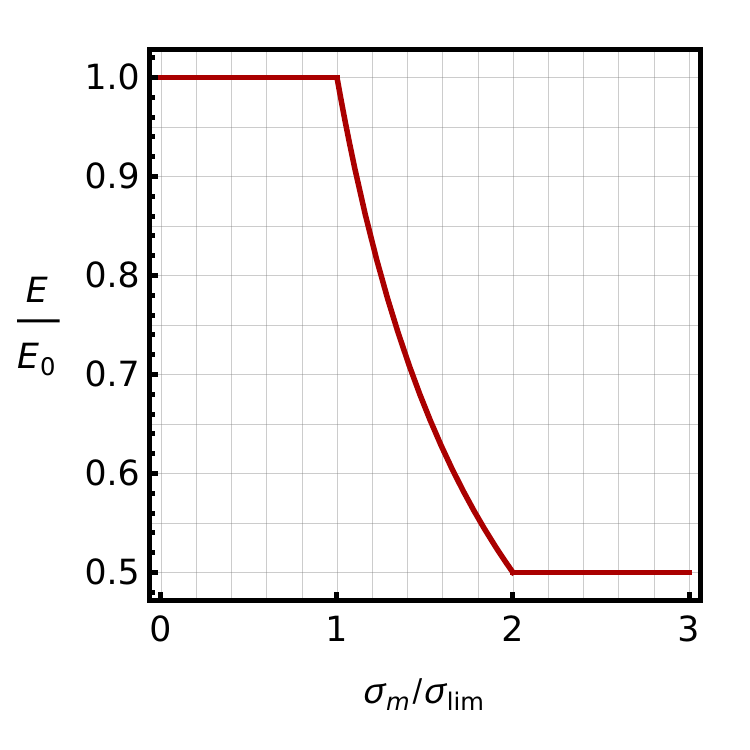}
    \caption{\Large (a)}
    \end{subfigure}
    \begin{subfigure}[a]{.45\linewidth}
    \centering
    \includegraphics[width=\linewidth]{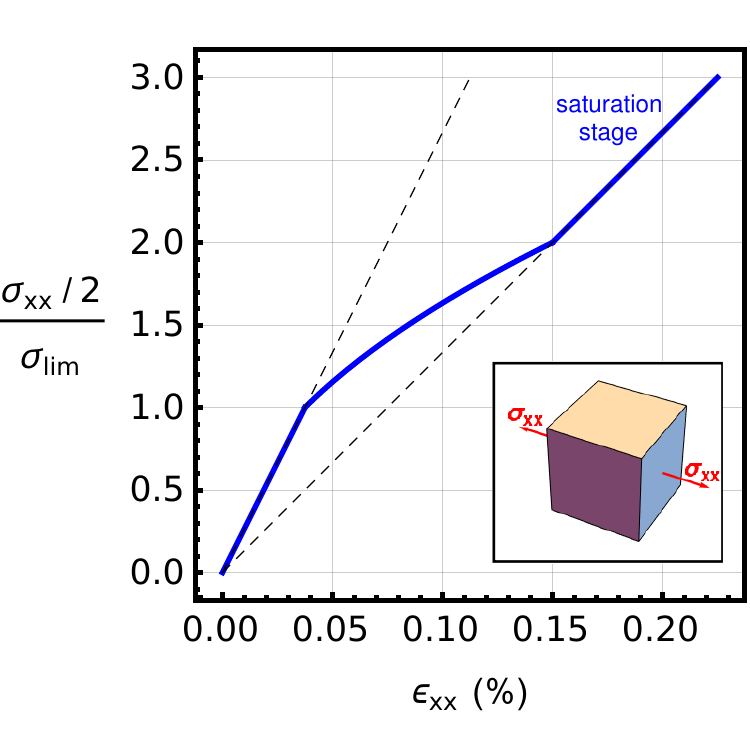}
    \caption{\Large (b)}
    \end{subfigure}
    \caption{Visualizing the material response of model \cref{eq:model_Y}. (a) Stiffness loss as mean stress increases. (b) Stress-strain relation for uniaxial loading (cf. Ref.~\cite{Ortiz:1987}).}
    \label{fig:material}
\end{figure}

\paragraph{Material} 
The material the plate is made of is assumed to change (decrease) stiffness when the mean stress ($\sigma_{m} = (\sigma_\text{I} + \sigma_{\text{II}} )/2 = (\sigma_\text{xx} + \sigma_{\text{yy}} )/2$) surpasses a limit value $\sigma_{\text{lim}}$. 
No inelastic behavior is considered in these simulations, deformations are considered recoverable; this does not play out anyway since we study a quasi-static monotonously-increasing loading scenario. 
The Young's modulus of the material is assumed to change according to an underlying model (\Cref{fig:material}):
\begin{align} \label{eq:model_Y}
    E (\sigma_m)
    =
    \begin{cases}
    E_0 \quad \text{if} \quad \sigma_m <  \sigma_{\text{lim}} \\
    \max \left[ \left( 
    { \sigma_{\text{lim}} \over \sigma_m } 
    \right) E_0 \, , \,
    0.5E_0
    \right] 
    \quad \text{if} \quad \sigma_m \ge  \sigma_{\text{lim}}
    \end{cases} \, ,
\end{align}
$E_0$ being the zero-strain modulus, while the Poisson's ratio is assumed to remain constant. For the simulations we are to show, $E_0 = 200 \, \mathrm{GPa}$, $\nu = 0.33$ and $\sigma_{\text{lim}} = 75 \, \mathrm{MPa}$. 
This model is ``invisible'' to the data-based solvers, but not for the NR one.
%
We use $\mathbb{C}_e = \mathbb{D}_e = \mathbb{D}$ $\forall e$, where
\begin{align}
    \mathbb{D}
    =
    {E_0 \over 1 - \nu^2}
    \begin{bmatrix}
    1 &\nu &0 \\
    \nu &1 &0 \\
    0 &0 &{{1 - \nu} \over 2 } 
    \end{bmatrix} \, .
\end{align}
We note that models of this type have been proposed as a means of quantifying the effect of crack shielding by microcracking in brittle materials such as ceramics \cite{Hutchinson:1987, Ortiz:1987}.


The mesh is generated with CST elements. The system is solved first with linear-elasticity elements to check that at least some elements, but not all, surpass the limit stress. A simulation where every element remains linear or becomes non-linear would not be an interesting case study for d-refinement.
After this first verification, we move to solve it also using pure DDCM, with d-refinement and with the Newton-Raphson (NR) solver.

At each loading level, the incremental NR solver convergence condition is $ |\boldsymbol{f}_{\text{ext}}-\boldsymbol{f}_{\text{int}}|/ |\boldsymbol{f}_{\text{ext}}|  < \mathrm{tol} $, where $\boldsymbol{f}_{\text{ext}}$ is the nodal external force vector and $\boldsymbol{f}_{\text{int}}$ the internal one for unrestrained degrees of freedom. The total external force is applied linearly over ten steps.

The datasets used for this first comparison proceed from the solution of the incremental solver: at the end of each loading step, the strain and stress state of each element is recorded as a point in phase space that becomes part of the material set $\mathrm{D}$ according to the criterion in \Cref{Sec:implementation_details}.

\subsubsection{Solution analysis}

Qualitatively, \Cref{fig:plate_mesh_results_b}, the method performs as expected: the top frame displays the elements that overcome the mean-stress limit in the linear-elastic simulation (vertically-hatched elements), while solid-color elements in the same figure show the spread of non-linearity predicted by the NR solution. 
The latter set is larger and contains most of the former. The stress is more widely distributed over the domain once the non-linear softer material is accounted for, what pulls more elements past the threshold than those presumed by the linear-elastic solver. 
The lower frame shows, in addition to the aforementioned NR elements that go non-linear, the elements that become data-driven according to d-refinement (horizontal hatching). See how these subsume completely the others, meaning that this solver does predict where non-linear behavior needs to be accounted for and proceeds in accordance to implement the DD formulation therein. 

\paragraph{Wall time speed-up} 
Regarding the results presented next, let us remark that the time difference between realizations of the same simulation differ by about than a $ \pm 10\%$ (maximum), in particular those solved with pure DDCM, in which the initial assignation is random. 

We begin by comparing the performance of the new approach against NR solver for different values of $\mathrm{tol}$ while raising the load up to its final value, $p = 100 \, \text{MPa}$, over ten increments at constant rate. Note that this final value would yield a state of mean stress $\sigma_m \approx 50\, \text{MPa} < \sigma_{\text{lim}}$ away from the hole.  

We take advantage of the most accurate NR simulation ($\mathrm{tol} = 10^{-5}$) to generate the dataset as described above and of the linear-elastic solution to set the method constants: $\mathbb{C}_e = \mathbb{D}$ $\forall e \in S_2$. 
%
The set $\mathrm{D}$ contains at first 29240 data (including information about elastic loading), and it reduces to just 886 (that only include information within the non-linear regime) after the sifting (\Cref{Sec:implementation_details}). 
Let us re-iterate that this convenient reduction is uncontroversial under the assumption of monotonously-increasing load, but new considerations enter the picture if one wants to consider non-monotonous loading (either elastic or inelastic); for instance, one would have to take decisions as to how to deal with DD elements that revert to the elastic regime. 

\begin{center}
\begin{tabular}{|P{2.cm}|P{2.cm}|P{2.cm}|P{2.cm}|P{2.cm}|P{2.cm}|}
%
%
\cline{2-6}
\multicolumn{1}{l|}{} 
& $\Delta t_{\text{d-ref}}$ (|D|=886)
& $\Delta t_{\text{NR}}$ ($\mathrm{tol} = 10^{-3}$) & $\Delta t_{\text{NR}}$ ($\mathrm{tol} = 10^{-4}$)
& $\Delta t_{\text{NR}}$ ($\mathrm{tol} = 10^{-5}$)
& $\Delta t_{\text{DD}}$ (|D|=4588)
\\ \specialrule{.1em}{.05em}{.05em}
\multicolumn{1}{|l|}{ $\qquad [s]$} 
& 20.3           
& 38.5
& 51.2 
& 72.8
& 928.7 \\
\multicolumn{1}{|l|}{increment } 
& -           
& $\times 1.9$         
& $\times 2.5$ 
& $\times 3.6$  
& $\times 45.8$
\\ \cline{1-6}
\end{tabular}
\captionof{table}{Solving the plane-stress problem with different methods: wall time comparison ($p = 100 \, \text{MPa}$, initialization type 1, 10 NR increments).}
\label{table:tab_plate_c-ref_1}
\end{center}


The results concerning computational speed-up are consigned in \Cref{table:tab_plate_c-ref_1}. 
The d-refinement technique is faster than both NR solver (for all the precision tolerances tried) and pure DD solution. For the latter, a simulation with a reduced material set ($|\mathrm{D}| = 4598$, including linear response) was sampled at random from the NR phase-space trajectories (29240 points). 
\begin{figure}
    \centering
    \begin{subfigure}[a]{.48\linewidth}
    \centering
    \includegraphics[width=\linewidth]{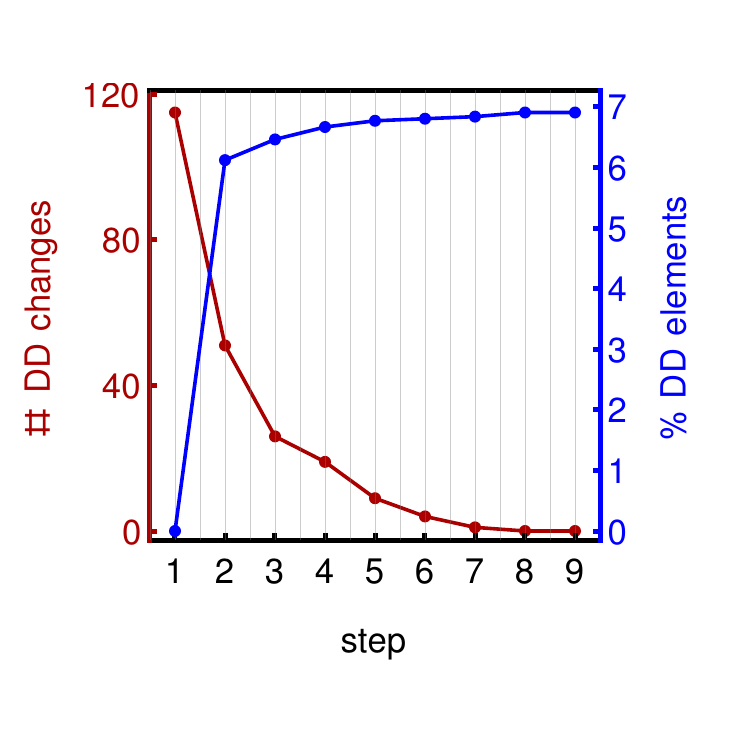}
    \caption{\Large (a)}
    \end{subfigure}
    \begin{subfigure}[a]{.48\linewidth}
    \centering
    \includegraphics[width=\linewidth]{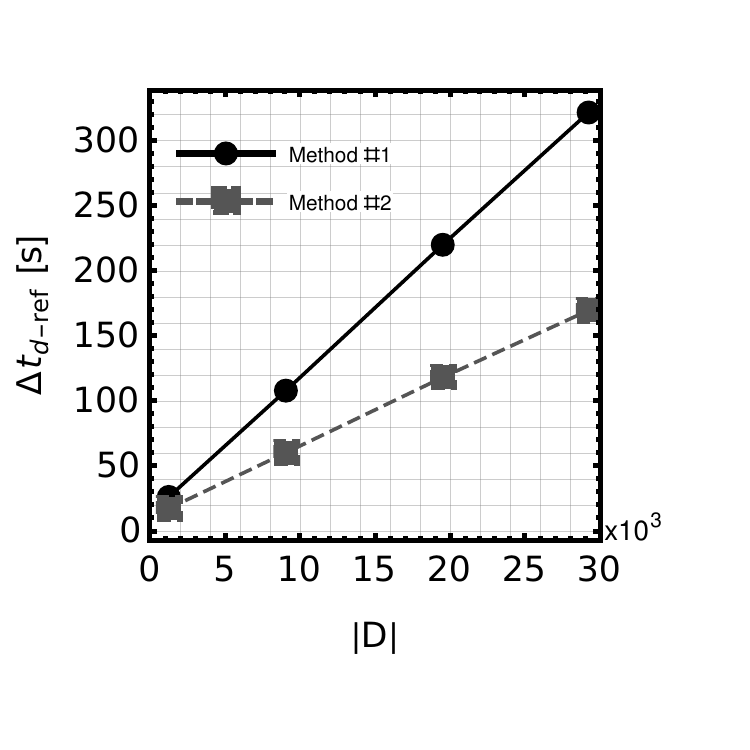}
    \caption{\Large (b)}
    \end{subfigure}
    \caption{(a) Simulation progress in terms of portion of the mesh being refined and number of DD elements that changed material data by the end of each step. (b) Linear scaling of d-refinement wall time ($\Delta t_{d-ref}$) with size of dataset (for a similar number of refined elements): the solid line (Method \#1) represents initial assignation $\boldsymbol{z}_e^* = \text{argmin}~d(\boldsymbol{z}_e^{FE} , \, \mathrm{D})$ (202 refined elements by the end of the simulation), the dashed line (Method \#2) $\boldsymbol{z}_e^* = \boldsymbol{0}$ (270 refined elements on average).}
    \label{fig:d-ref_1_evolution}
\end{figure}


The evolution of this particular d-refinement simulation is represented in \Cref{fig:d-ref_1_evolution}. All the elements that were presumed by the linear-elastic solver to be above the mean-stress threshold were turned into DD after the first iteration. Of course, all these elements were assigned new dataset points, what is reflected in the other (red) curve first point. As the simulation moved forth, the number of DD elements converged quickly, while the material point assigned to each one of them did so more slowly. 

\paragraph{Accuracy of d-refinement} 
To assess the difference between the solution using mesh d-refinement and the most accurate incremental solver ($\text{tol} = 10^{-5}$), we choose to compare the phase-space distance between the points corresponding to the Newton-Raphson solution ($\boldsymbol{z}^{\text{NR}} \in \mathbb{R}^{N_{e} \times 2 N_c}$, $N_c = 3$ for plane stress) and the ones of the d-refinement solution ($\boldsymbol{z}^{\text{d-ref}} \in \mathbb{R}^{N_{e} \times 2 N_c}$):
\begin{align} \label{eq:ratio_distances}
    { |\boldsymbol{z}^{\text{d-ref}}
    -
    \boldsymbol{z}^{\text{NR}}|
    \over 
    |\boldsymbol{z}^{\text{d-ref}}|
    }
    =
    {
    \left(
    \sum_{e=1}^{N_e} 
    w_e |\boldsymbol{z}_e^{\text{d-ref}}
    -
    \boldsymbol{z}_e^{\text{NR}}|
    \right)^{1/2}
    \over
    \left(\sum_{e=1}^{N_e} 
    w_e |\boldsymbol{z}_e^{\text{NR}}|
    \right)^{1/2}
    } 
    \approx 0.034
    \, ,
\end{align}
that is, the distance between solutions in phase space is less than $4 \%$ of the one from the origin to the d-refinement solution. 
This is in spite of the material set containing exactly the same points as the non-linear portion of the Newton-Raphson trajectory. 
Actually, if we compare where the refined elements land, only 23 out of 202 arrive at their final configuration in the incremental solver. 
This does not translate in error nonetheless (recall that the error we have estimated using distances is less than a $4\%$), which necessarily means that the local density of the material set in phase space is such that some elements can converge to nearby points at virtually no cost. 

For this kind of initialization ($\boldsymbol{z}_e^* = P_D \boldsymbol{z}^{FE}$, labeled ``method \#1''), it is interesting to note that the proportion of elements that become data-driven is either completely independent of or very weakly-dependent on the phase-space density. 
Increasing the precision of the incremental solver by raising the number of load steps from 10 to 25 increases the material dataset size to $|\mathrm{D}| = 1878$ points, and this has no effect either over the number of elements that become DD (202) or on the proportion of these that reach the same destination (23). 
In terms of computational time, \Cref{table:tab_plate_c-ref_2}, d-refinement again outperforms the most accurate NR ($\mathrm{tol} = 10^{-5}$).

\begin{center}
\begin{tabular}{|P{2.cm}|P{2.cm}|P{2.cm}|}
%
%
\cline{2-3}
\multicolumn{1}{l|}{} 
& $\Delta t_{\text{d-ref}}$ (|D|=1858) 
& $\Delta t_{\text{NR}}$ ($\mathrm{tol} = 10^{-5}$) 
\\ \specialrule{.1em}{.05em}{.05em}
\multicolumn{1}{|l|}{ $\qquad [s]$} 
& 55.3           
& 122.3 \\
\multicolumn{1}{|l|}{increment } 
& -           
& $\times 2.2$         
\\ \cline{1-3}
\end{tabular}
\captionof{table}{Solving the plane-stress problem: wall time comparison ($p = 100 \, \text{MPa}$, initialization type 1, 25 NR increments).}
\label{table:tab_plate_c-ref_2}
\end{center}

\paragraph{Effect of dataset size on d-refinement performance} 
This also motivates a sensitivity study as to the influence of the size of the material set D (number of material data): 
the incremental solver is kept at ten load increments and $\mathrm{tol} = 10^{-3}$, but we grow $\mathrm{D}$ by lowering the threshold to facilitate points in the Newton-Raphson trajectory entering D. 
%
%
%
We find the expected linear proportionality between total wall time and dataset size, see solid line in \Cref{fig:d-ref_1_evolution}. We discuss the influence of the other initial data assignation (which happens to reduce time, see dashed line in \Cref{fig:d-ref_1_evolution}, at the expense of accuracy) in the next section, but both methods display this linear scaling. 
To further decrease the computational time, one could use some of the techniques discussed in the literature \cite{EGGERSMANN_2,tensor_voting}.


\paragraph{Influence of initial data assignation}

Until now, reported results used phase-space searches to assign an initial datum to each element that evolves from FEM to DD. This involves additional searches that could be avoided by always assigning a pre-defined initial datum, i.e., the origin.  

This creates a somewhat disjoint material set $\mathrm{D}$, since there is a single point away from the cluster that represents non-linear behavior. 

The approach brings about a substantial speed improvement: based on the results presented in \Cref{fig:d-ref_1_evolution}, we find that this initial assignation method (``method \#2'') reduces the computation time by about 45\%. This comes at the expense of increasing the error when compared to NR results: the distance ratio \cref{eq:ratio_distances} doubled from 3.5\% to 7\%. 
So, as often, there is here a trade-off between accuracy and time efficiency. Depending on the priority, one would favor one method over the other. However, this conclusion is contingent on localized non-linear behavior, what leads to having similar final number of DD elements, see next point.

\paragraph{Impact of spreading non-linear behavior} 
As the load increases, the actual solution departs more meaningfully from the linear-elastic one. 
As elements excursion deeper past the threshold, richer datasets are also necessary to capture the response. 

For instance, keeping the dataset inclusion criterion and increasing by 20\% the load we have been using leads to an increase in $|\mathrm{D}|$ from 886 to 3419 phase-space points.

This also increases both the NR computation time and the d-refinement's, as well as the relative error between them (\cref{eq:ratio_distances}), see \Cref{table:tab_plate_c-ref_3}. The spreading of the non-linearity induced by the intenser load means more DD elements and hence more phase-space searches, what slows d-refinement down to the point of NR being faster in this case.   
The final number of refined elements has escalated from 202 to 620. Interestingly, initialization type 2 ($\boldsymbol{z}_e^* = \boldsymbol{0}$) ends up defaulting to a pure data-driven mesh due to accumulation of errors and the total time skyrockets. 

\begin{center}
\begin{tabular}{|P{2.cm}|P{2cm}|P{2cm}|P{3.cm}|}
%
%
\cline{2-4}
\multicolumn{1}{l|}{} 
& \multicolumn{2}{l|}{$\qquad \Delta t_{\text{d-ref}}$ (|D|=3419)} 
& \multirow{2}{*}{\parbox{2cm}{\centering $\Delta t_{\text{NR}}$ ($\mathrm{tol} = 10^{-5}$)}}\\
\cline{2-3}
\multicolumn{1}{l|}{} 
& Method \#1 
& Method \#2
& \\
 \specialrule{.1em}{.05em}{.05em}
\multicolumn{1}{|l|}{ $\quad \qquad [s]$} 
& 154.8
& 924.3
& 105.3 \\
\multicolumn{1}{|l|}{$\quad$increment } 
& $\times 1.5$ 
& $\times 8.8$
& -  \\ 
\cline{1-4}
\multicolumn{1}{|l|}{error (\cref{eq:ratio_distances})}
& $0.05$ 
& $ 0.14$
& -  \\ 
\cline{1-4}
\end{tabular}
\captionof{table}{Solving the plane-stress problem: wall time comparison and accuracy ($p = 120 \, \text{MPa}$, different initialization types, 10 NR increments). The NR solution is used as reference to compare in this case.}
\label{table:tab_plate_c-ref_3}
\end{center}

We remarked that the precision of either approach can be increased by sharpening the refinement threshold criterion.
%
In any case, such an approach is a barren one, since in practice the actual threshold will not be well-defined; the difference between d-refinement and the incremental solver could be regarded not as a demerit of the former but as modeling bias infused in the latter.

The main conclusion is that in quasi-static, monotonous-loading cases where non-linearity is expected to be present well over the domain, d-refinement must be used with initialization \#1, since this yields better first guesses, thus boosting convergence and limiting the number of extra elements to be refined. 
To gauge the extent of the non-linear portion beforehand, comparing the results of a linear-elastic calculation to the threshold may suffice. We have shown that initialization method \#2 is adequate and efficient when there is localized non-linearity and, in the next section, we will also prove its suitability when used in tandem with appropriate load increments.



\subsection{1D bar elements in 3D space: octet-truss structure} 
\label{sec:verification_truss}

In this study, we create a dataset by sampling a constitutive law, and compare d-refinement and NR solver
(a) changing the density of the dataset, and (b) changing the number of loading steps.

This application is implemented in a Jupyter notebook, which is provided as supplementary material.

\subsubsection{System}
\begin{figure}
    \centering
    \includegraphics[width=0.75\textwidth]{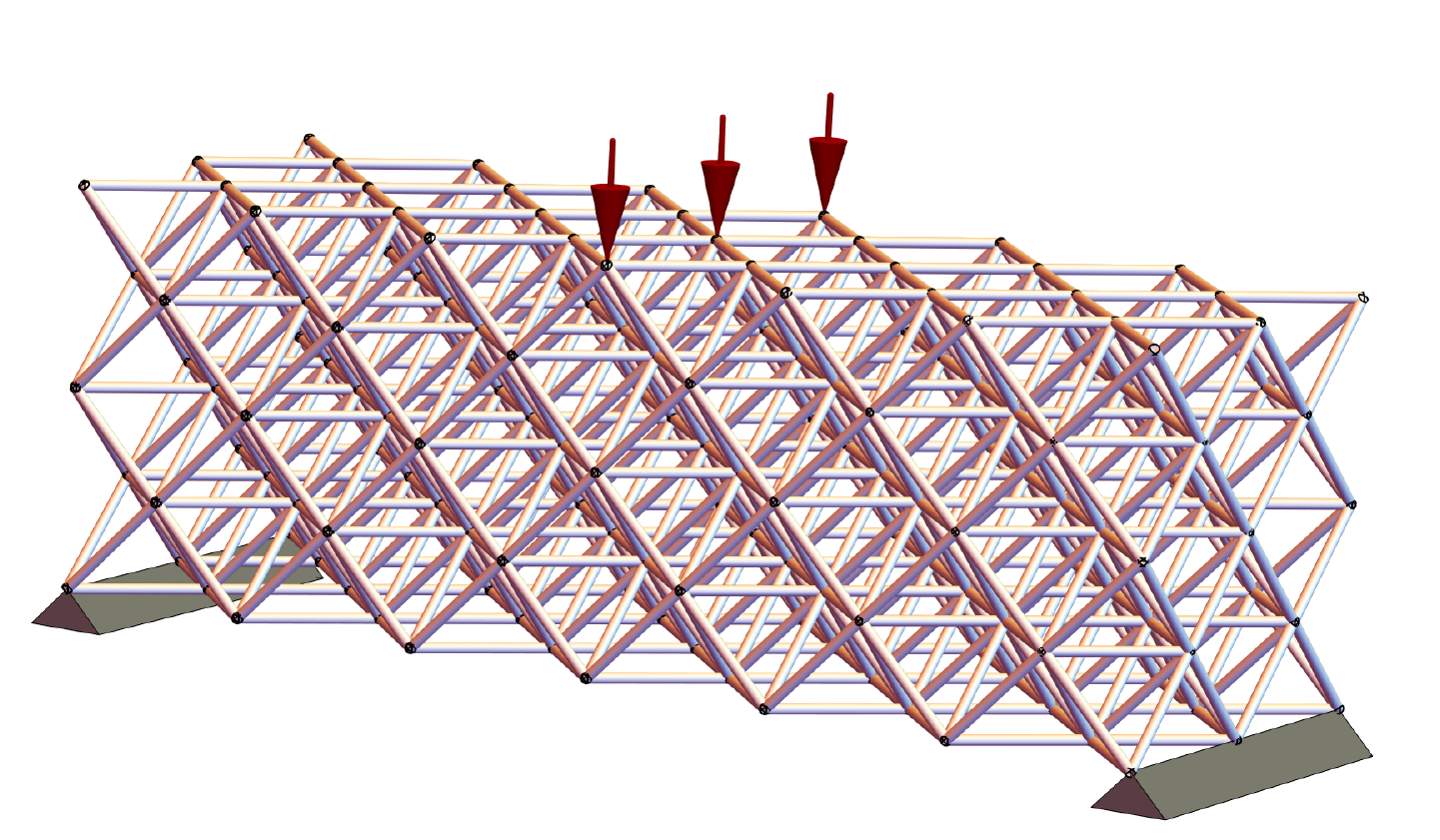}

    \caption{Scheme of three-point bending configuration. Arrows represent the imposed displacement at the top middle nodes}
    \label{fig:octet_truss}
\end{figure}

The second study features a truss made of octet unit cells \cite{Fleck-Deshpande}. The model is made up by axially-loaded bar elements (no bending), with linearized kinematics, and a total of $6 \times 2 \times 2$ unit octet truss cells submitted to a 3-point bending test as represented in \Cref{fig:octet_truss}. The parameters of the unit cell octet truss were inspired by the work in Ref.~\cite{toughness_metameterials} and are summarized in Table \ref{table:truss_parameters}. It should be noted that the goal is not to reproduce their results: to do so, either buckling in the case of slender member or joint stiffness for thick member should have been accounted for. This model is intended to test and explore the performance of the d-refinement solver. It has two main advantages: bar elements are straightforward to discretize and their local phase-space is two-dimensional: $\boldsymbol{z}_e= (\epsilon_e, \sigma_e) \in \mathbb{R}^{2}$.

The beam displacements are fixed on both lower edges and a displacement-controlled deflection is imposed on the top middle nodes. 
While the load is applied in steps, the problem is considered to be quasi-static and thus solved with the static solver presented above. See \Cref{fig:octet_truss} for a graphical representation. 

\begin{figure}
    \centering
    \includegraphics[width=0.75\textwidth]{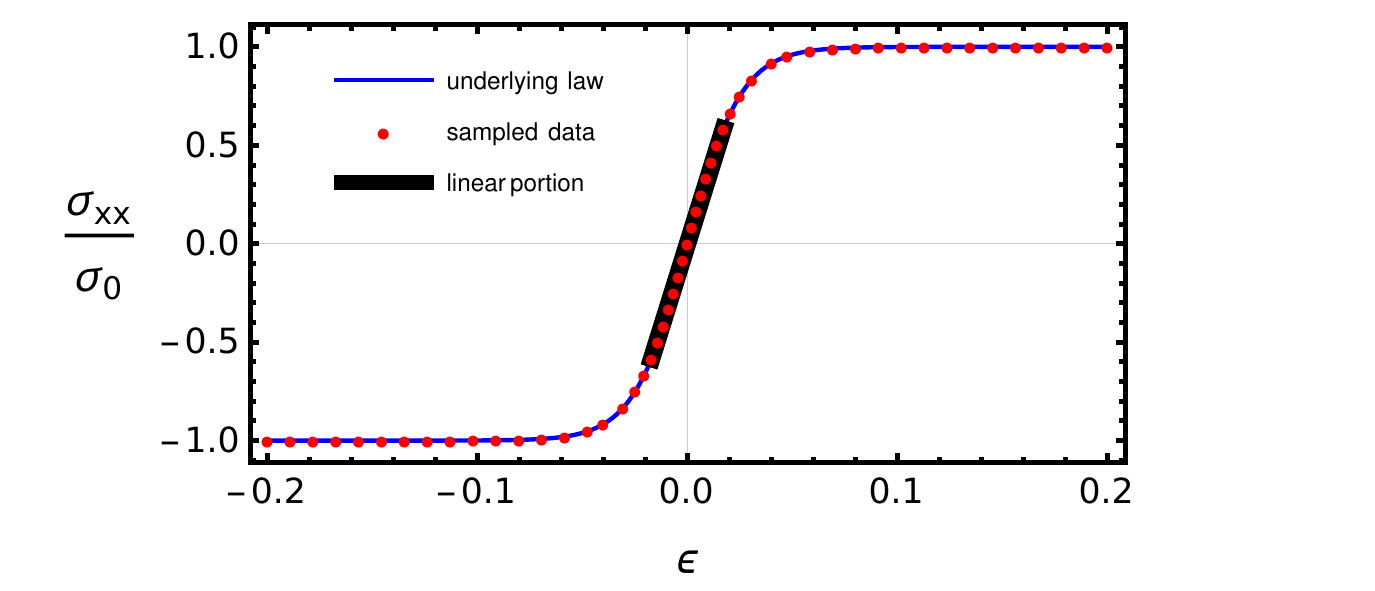}
    \caption{Constitutive law, synthetic dataset and the limits of the linear elastic domain. For the dataset, only 1\% of the points are plotted to increase legibility. }
    \label{fig:dataset_1D}
\end{figure}

\paragraph{Material} The behavior of the bulk material is also inspired from Ref.~\cite{toughness_metameterials}, in which trimethylolpropane triacrylate (TMPTA) is used. The material displays similar Young's modulus at zero-strain $E_0=430 \, \text{MPa}$ and limit strength $\sigma_f=11 \, \text{MPa}$. However, this artificial material is assumed to be non-linear elastic, as opposed to brittle, and is represented by a hyperbolic-tangent stress-strain relation (see \Cref{fig:dataset_1D}):
\begin{align}
    E(\epsilon) 
    &= 
    E_0 
    \left[1- \tanh^2
    \left(
    \frac{E_0}{\sigma_l} \epsilon 
    \right) \right], \\
    \sigma(\epsilon) 
    &= 
    \sigma_f \, 
    \tanh
    \left(
    \frac{E_0}{\sigma_l}  
    \epsilon
    \right) \, .
\end{align}


A synthetic database of 2700 points is generated by sampling uniformly the constitutive law between $\epsilon = -0.2$ and $\epsilon = 0.2$. 
For the purpose of d-refinement, the trigger to leave the linear elastic domain is assumed to be $\sigma > \sigma_{l} = 7 \, \mathrm{MPa}$. \Cref{fig:dataset_1D} summarizes the material behavior, with the underlying constitutive law, the dataset and the assumed linear elastic domain.

\begin{table}[H]
    \centering
    \begin{tabular}{ccl}
        Notation & Value & \\
        \hline
        $l$ & $530 \, \mu $m & Strut length \\
        $d$ & $65 \, \mu $m & Strut diameter \\
        \hline
        $E_0$ & $430 \, \text{MPa}$ &  TMPTA Young's modulus at zero strain \\
        $\sigma_f$ & $ 11 \, \text{MPa}$ & TMPTA tensile strength \\
    \end{tabular}
    \caption{Summary of the different parameters used in the truss models with both the geometric parameters and the bulk material properties.}
    \label{table:truss_parameters}
\end{table}


\subsubsection{Solution analysis}

The following is a comparative study of the results obtained with three different solvers: a Newton-Raphson informed by the constitutive law (NR), a pure data-driven one informed by the dataset from the constitutive law (DDCM) and a d-refinement one informed by the dataset as well as the zero-strain Young's modulus and linear elastic limit (DR). Initialization method \#2 is chosen in this case. Unlike the previous plane-stress study, the incremental version of d-refinement is used here.

\paragraph{Error analysis}
The load response of the beam for each solver is recorded to be then compared. With a 2700-point dataset ($|\mathrm{D}| = 2700$) and taking the Newton-Raphson solution as a reference, the relative errors in load response are 2.8\% and 1.8\% for the DD and DR solvers, respectively. 
\paragraph{Convergence with noiseless and noisy dataset} Then, the number of points in the dataset is increased. The errors, as defined by \cref{eq:ratio_distances}, of both the DD and DR solvers reduce at a rate given by a power law with exponent one. This $\sim |\mathrm{D}|^{-1}$ scaling is coherent with the literature \cite{Trent_1}. 
However, for the DR solver, the convergence to the NR solution is ultimately limited by the quality of the linear assumption at small strains. The linear approximation does not overlap perfectly with the underlying hyperbolic-tangent law, hence some minor difference in the linear-elastic elements persists independently of the number of datapoints. 
We stress that this is not a flaw in the d-refinement method but a matter of modeling bias.

When random Gaussian noise is added to the datapoints (both in strain and stress) with standard deviation inversely proportional to the square root of the number of datapoints \cite{Trent_1}, the same behavior is observed but the convergence rate slows down to $|D|^{-1/2}$ which is again in agreement with the literature results \cite{Trent_2, Trent_3}. For data whose noise is independent of the number of datapoints, max-ent data-driven solvers \cite{Trent_3} are more adequate than the fixed-point one presented here.
As for a more formal proof of convergence, the linear elastic approximation can be seen as being part of the dataset (as a graph or a sampling of a line with infinite density for example) and thus the convergence proof derived in Ref.~\cite{Conti_Mueller_Ortiz} can be extended to the d-refinement solver: if the linear approximation and the dataset converge to the real underlying material behavior, so will do the d-refinement solver.


\begin{figure}[H]
    \centering
    \includegraphics[width=0.75\textwidth]{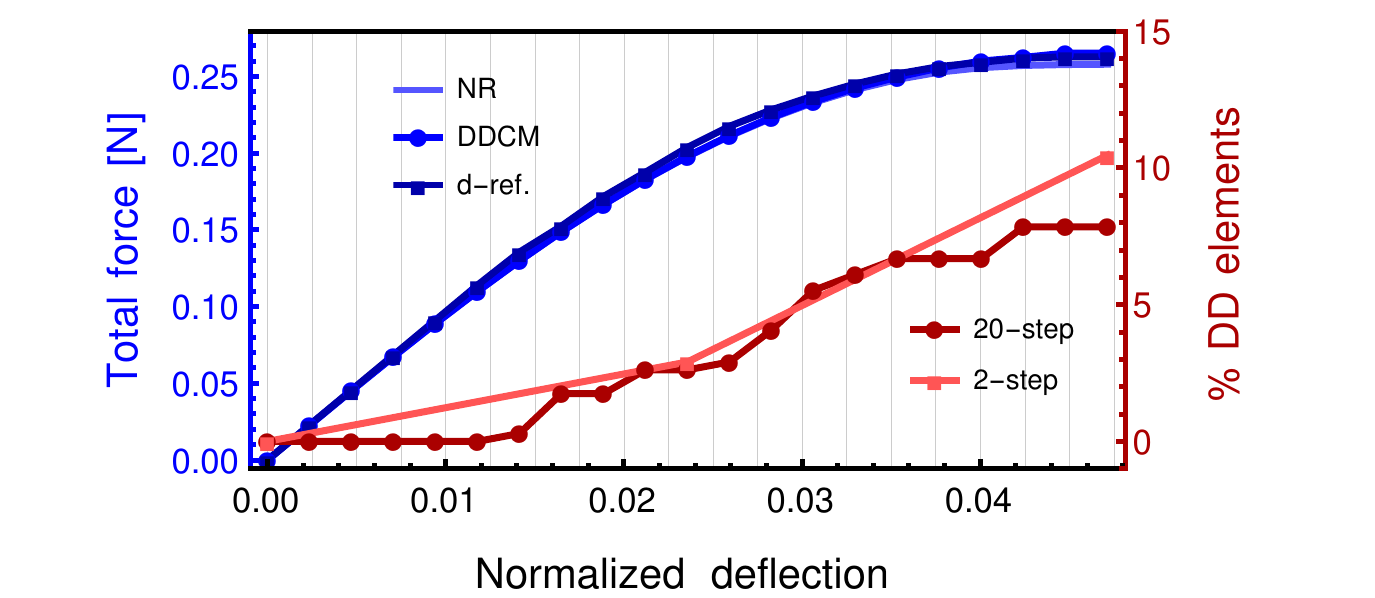}
    \caption{Load-deflection curves (upper) obtained with the three solvers and a 20-step loading protocol (left vertical axis). The deflection is normalized by the beam length.
    The total force is computed through the internal forces in the bars that converge at the nodes where the displacement is enforced.
    The lower curves (right vertical axis) represent the fraction of rod elements that have been refined, i.e. switched to DD, for both the 20-step loading and the 2-step loading.}
    \label{fig:load_deflecion}
\end{figure}

\begin{figure}
    \centering
    \captionsetup{labelformat=empty}
    \begin{subfigure}[a]{.45\linewidth}
    \includegraphics[width=\linewidth]{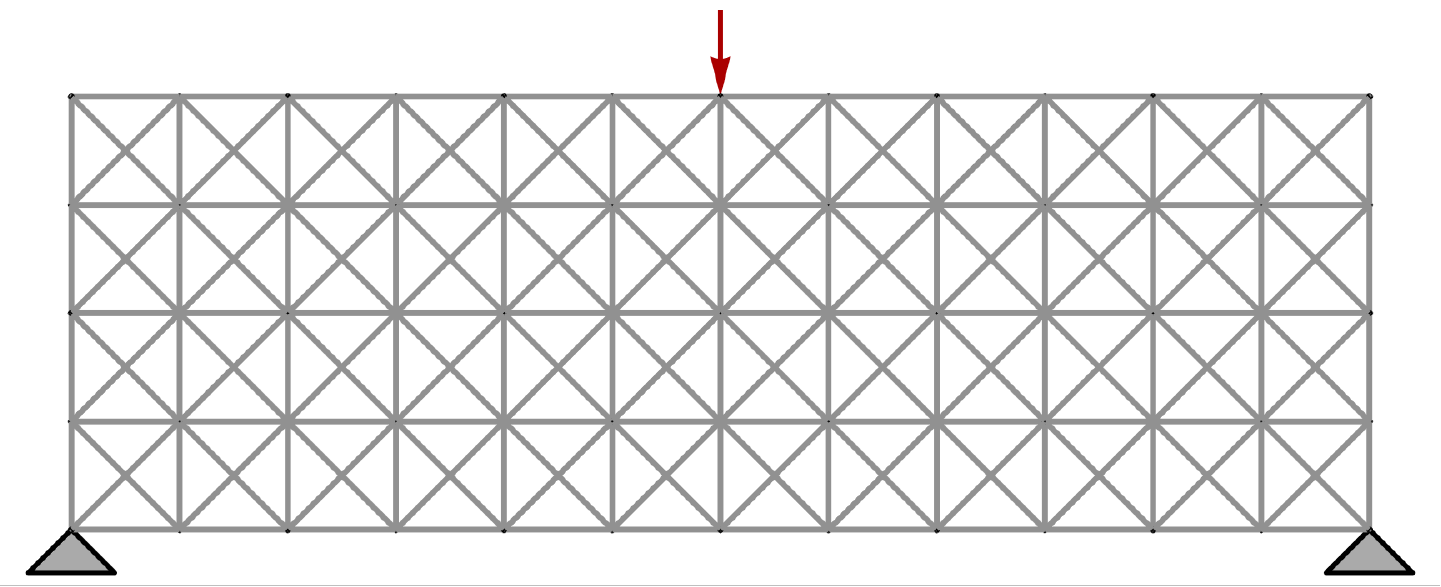}
    \caption{(a) 0\% load}
    \vspace{0.5cm}
    \end{subfigure}
    
    \begin{subfigure}[a]{.45\linewidth}
    \includegraphics[width=\linewidth]{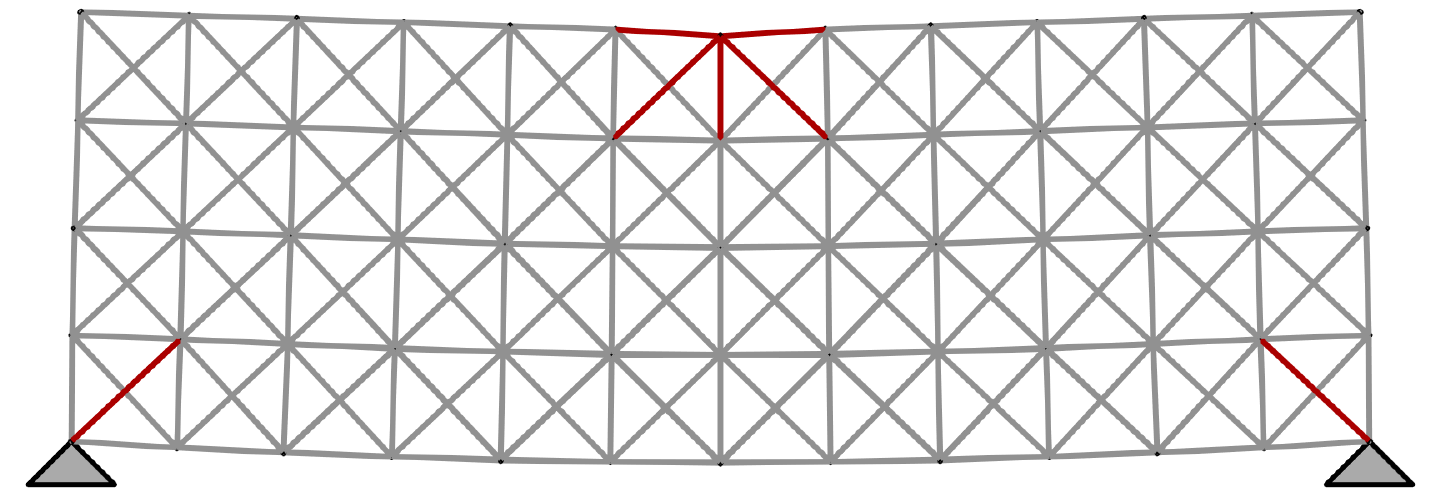}
    \caption{(b) 20 steps -- 50\%load}
    \includegraphics[width=\linewidth]{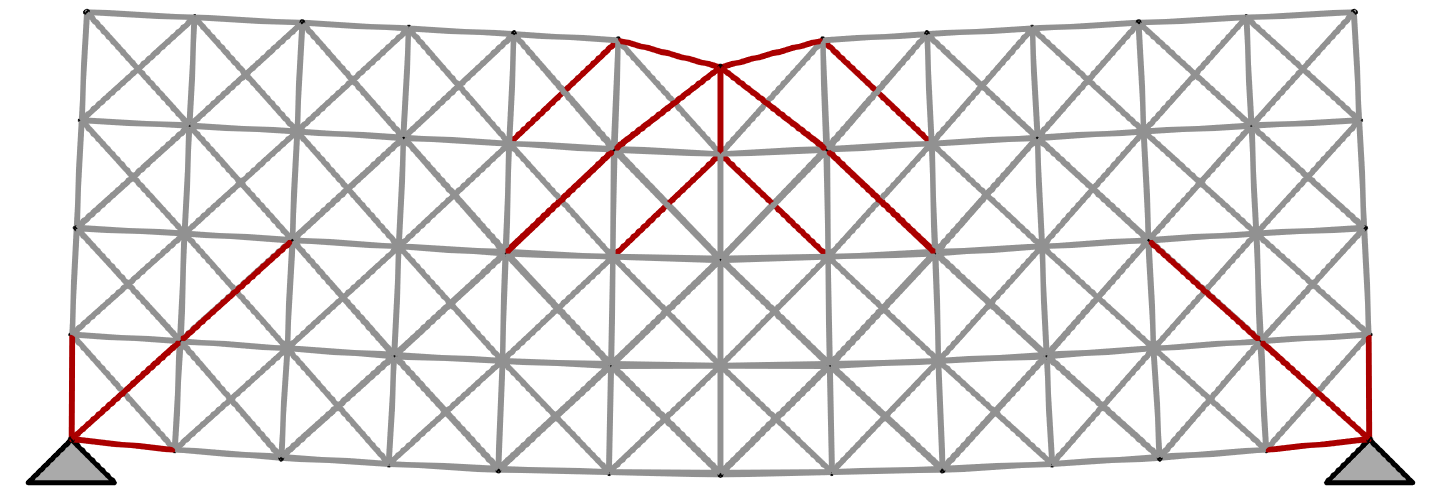}
    \caption{(c) 20 steps -- 100\%load}
    \end{subfigure}
    \begin{subfigure}[a]{.45\linewidth}
    \includegraphics[width=\linewidth]{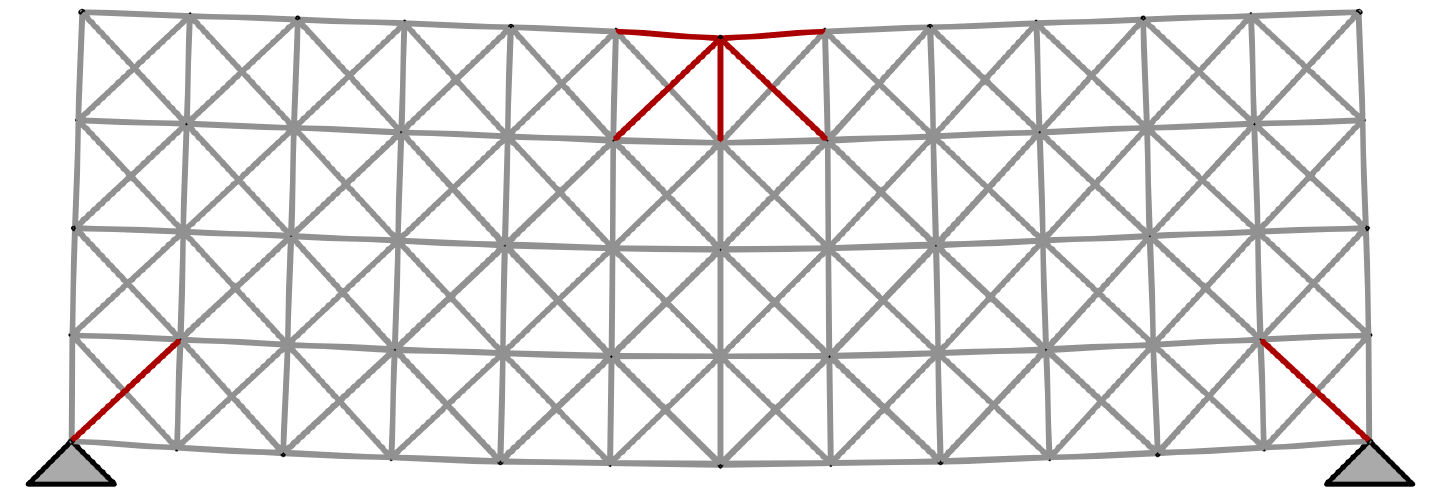}
    \caption{(d) 2 steps -- 50\%load}
    \includegraphics[width=\linewidth]{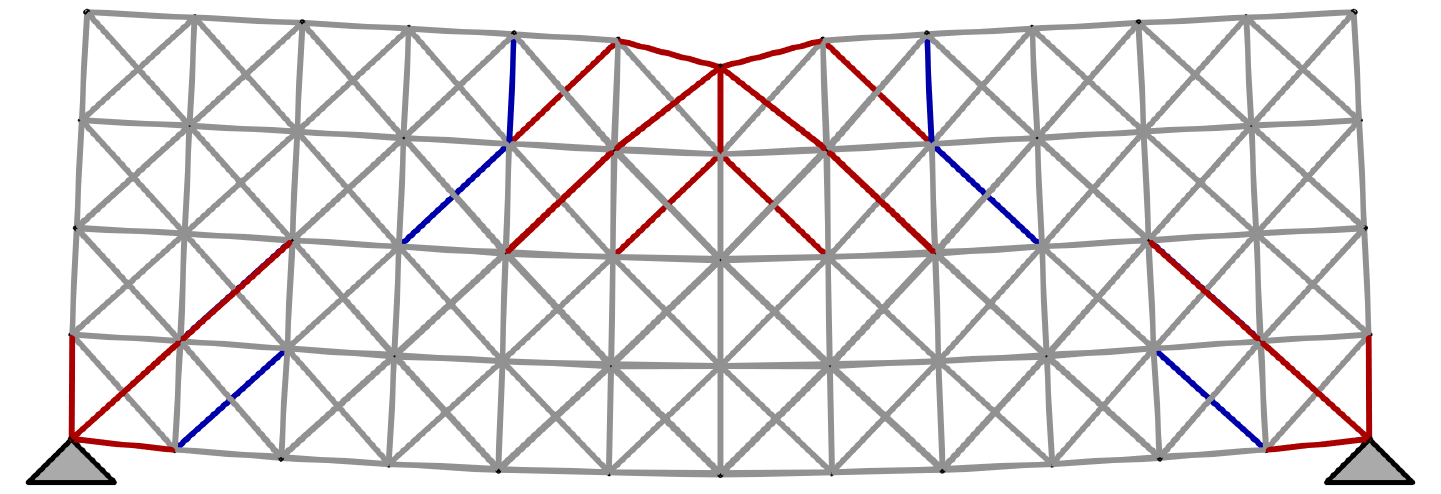}
    \caption{(e) 2 steps -- 100\%load}
    \end{subfigure}
    \captionsetup{labelformat=original}
    \caption{Projection of the beam truss on a 2D plane. The right column depicts results obtained with a 2-step loading and the left one with a 20-step loading. In red, elements that should be and are refined. In blue, elements that are unnecessarily refined.}
    \label{fig:c_ref_truss}
\end{figure}

\paragraph{Influence of load-stepping} As the stepping procedure unfolds and the load increases, more elements go over the threshold and are switched to the DD solver. 
The size of the loading step influences which elements are switched: a large step may cause the refinement to overshoot, i.e., to refine elements that should have remained within the linear elastic regime. The red lines (lower) in \Cref{fig:load_deflecion} highlight this phenomenon: the fraction of refined elements at the end of the loading is 7\% with 20 steps and 10\% with 2 steps. In \Cref{fig:c_ref_truss}, with 20-step loading, the refined elements correspond exactly to the ones that undergo large straining according to the NR solution. In 2-step loading, more elements are refined. 
This overshooting, associated to using initialization at the origin and not enough load steps, does not influence the accuracy of the final solution provided that 
(a) the dataset has points in the elastic zone, at least in the proximity of the limit as suggested before, and 
(b) the refined elements represent a fraction of the total (recall the results obtained in the previous case study when increasing the load). 
The only cost of the overshooting in refinement is computational, no meaningful extra errors are induced. It should be noted that the standard Newton-Raphson solver could diverge if the load steps were too large, thus another advantage of the d-refinement method is its stability.

\section{Application: bridging scales to study the fracture process zone in architected metamaterials} \label{Sec:application}

\begin{figure}
    \centering
    \includegraphics[width=\textwidth]{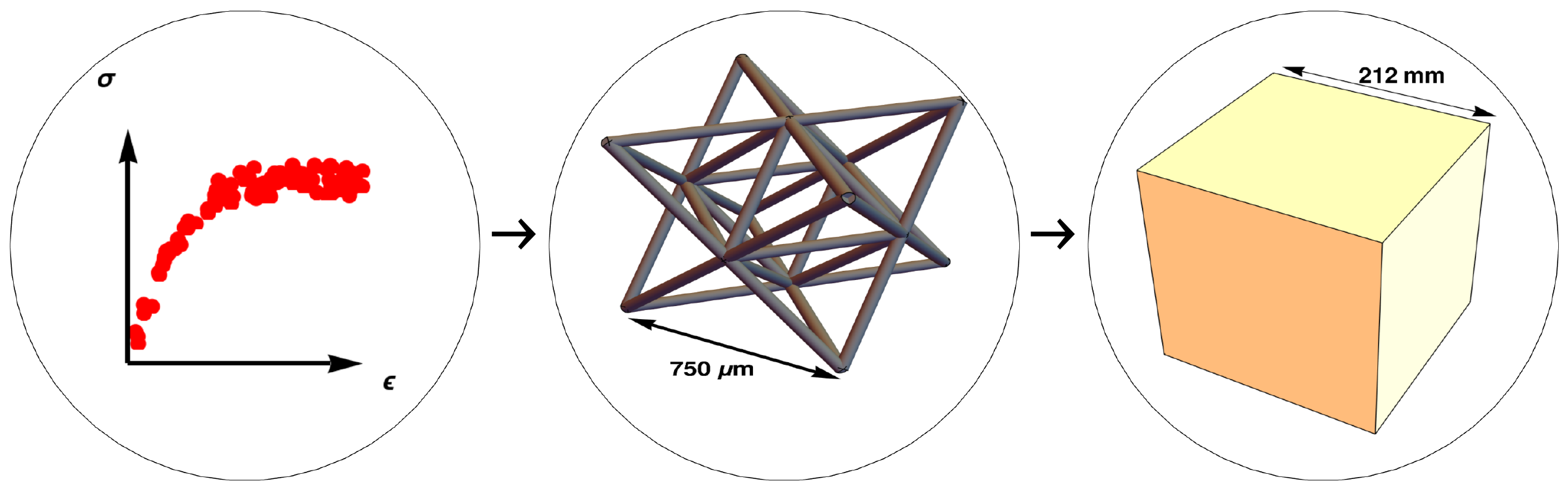}
    \caption{Scheme of the scale bridging procedure (from left to right): rod material dataset is used to characterize the response of octet-truss unit cell using DDCM, these RVE simulations yield linear-elastic constants and a non-linear material-response dataset used for macroscale simulations of a cubic sample of architected material using d-refinement.}
    \label{fig:bridging}
\end{figure}

While DDCM allows to use experimental data to make predictions without resorting to phenomenological models, it can also be applied as an efficient method to solve multiscale problems \cite{Kostas_1,Kostas_2,DD_foam}. It is particularly suited in the case where a convincing microscale model exists, which, for any reason, can not be run at the larger scale of interest. In such a case, a dataset consisting of RVE microscale simulations over different loading paths can be computed offline and then passed to a macroscale DDCM model. 
Since the overhead associated to running microscale simulations is much lower than the one of performing experiments, the development of the dataset can be done in several iterations: areas of the phase space which have been found to not be dense enough 
can be repopulated \cite{Kostas_1}; one can even foresee launching the RVE microscale simulations on the fly \cite{Kostas_1}.

To showcase the merits of DDCM for multiscale modeling, the two models presented in Section \ref{Sec:validation} are combined to devise an architected metamaterial \cite{Meza}. Starting from the synthetic dataset of TMPTA material behavior, microscale simulations of an RVE unit cell of octet truss were run using the DDCM solver. Then, this dataset was fed to a continuous macroscale model with the goal of resolving the stress distribution around a crack tip with the d-refinement solver. 

In summary, we will go from a dataset of the underlying material to the response of the microstructure, to finally reach the macroscale and perform mechanical analysis where the non-linear behavior of the octet-truss architected metamaterial is taken into consideration with no recourse to any constitutive modeling whatsoever.

This family of artificial materials has commanded the attention of the research community in recent decades \cite{Deshpande_Fleck_2}. Due to their manufacturability and remarkable properties, they are poised to challenge conventional materials in a bevy of applications, from thermal insulation \cite{thermal} to impact absorption \cite{Portela}. 

\subsection{System}

\begin{figure}[H]
\centering
\captionsetup[subfigure]{justification=centering}
\begin{subfigure}[b]{.32\linewidth}
\includegraphics[width=\linewidth]{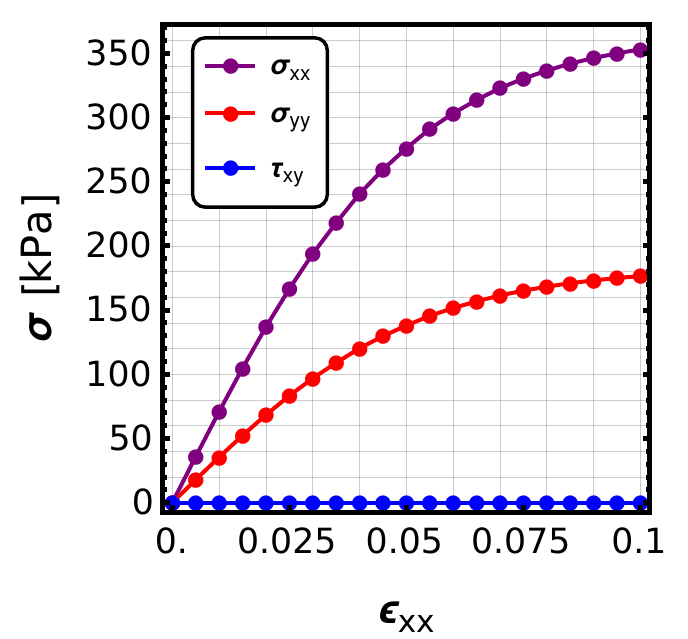}
\caption{\Large (a)}\label{fig:1a}
\end{subfigure}
\begin{subfigure}[b]{.32\linewidth}
\includegraphics[width=\linewidth]{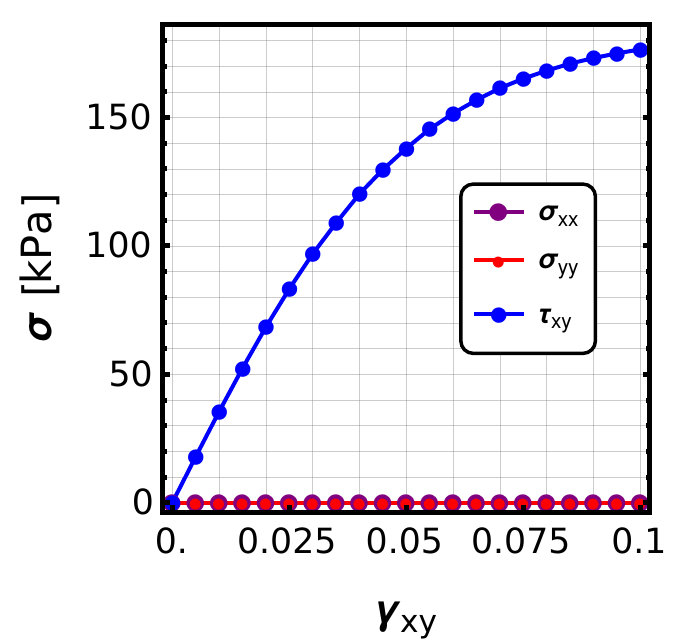}
\caption{\Large (b)}\label{fig:1b}
\end{subfigure}
\begin{subfigure}[b]{.32\linewidth}
\includegraphics[width=\linewidth]{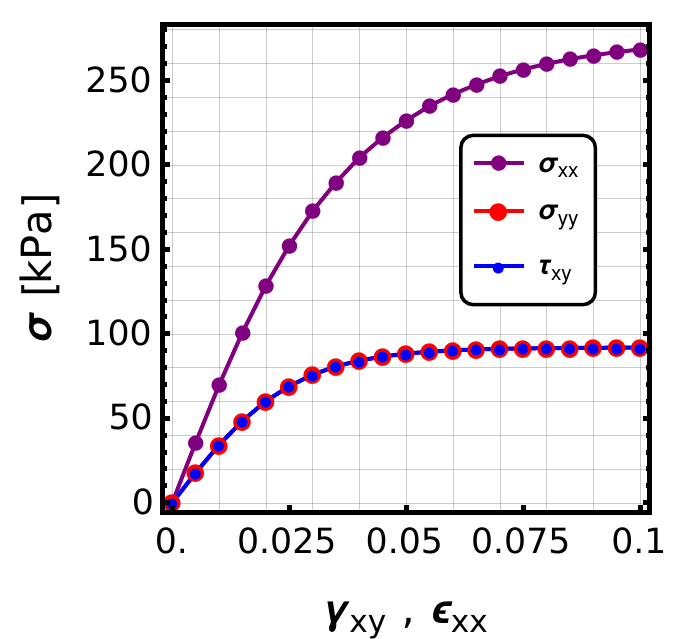}
\caption{\Large (c)}\label{fig:1c}
\end{subfigure}
    \caption{Strain-stress paths of microscale RVE simulations. %
    (a) Pure uniaxial elongation.
    (b) Pure shear.
    (c) Simultaneous elongation and shear.
    }
    \label{fig:strainstressRVE}
\end{figure}

\paragraph{Microscale} The microscale model is the same used in \Cref{sec:verification_truss}. 
The RVE is a single unit octet-truss cell (represented in the middle of \Cref{fig:bridging}). 
The axes of the frame of reference ($\boldsymbol{x},\boldsymbol{y},\boldsymbol{z}$) are aligned with the lattice planes ([100], [010], [001]), see \Cref{fig:1a}. The cross-sections of the bar elements that lie on the boundary are reduced as they are shared with the neighboring cells. 

To create the dataset, the stress response to specified strain states ($\epsilon_{xx}\,\epsilon_{yy}\,\epsilon_{xy}\,$) are simulated. For this, the strain state is converted into imposed displacements at the ten boundary nodes. 
To compute the homogenized stress states, the sum of the forces, on each facet, for each direction, is divided by the apparent area of the cell facet. 
The material behavior is again considered elastic. Since it was verified previously that the accuracy of the final solution does not depend on the number of loading steps when using initialization method \#2, no incremental loading is employed. The material response of the unit cell is illustrated in Figure \ref{fig:strainstressRVE} where strain-stress paths are plotted for uni-axial elongation, pure shear and combined shear-stretching.

\paragraph{Unit-cell elastic response} First, as the d-refinement method necessitates information about the small-strain behavior, the elastic constants should be determined. 
For this, the small-strain response of the unit cell to different deformation scenarios was solved with DDCM. The following compliance matrix is obtained:
\begin{align}
    \begin{pmatrix}
        \epsilon_{xx} \\
        \epsilon_{yy} \\
        \epsilon_{zz} \\
        \gamma_{yz} \\
        \gamma_{xz} \\
        \gamma_{xy}
    \end{pmatrix} = \frac{1}{0.1E_0}
    \begin{pmatrix}
        9 & -3 & -3 & 0 & 0 & 0 \\
        -3 & 9 & -3 & 0 & 0 & 0 \\
        -3 & -3 & 9 & 0 & 0 & 0 \\
        0 & 0 & 0 & 12 & 0 & 0 \\
        0 & 0 & 0 & 0 & 12 & 0 \\
        0 & 0 & 0 & 0 & 0 & 12 \\
    \end{pmatrix} 
    \begin{pmatrix}
        \sigma_{xx} \\
        \sigma_{yy} \\
        \sigma_{zz} \\
        \sigma_{yz} \\
        \sigma_{xz} \\
        \sigma_{xy}
    \end{pmatrix},
\end{align}

with $\gamma_{ij} = 2\epsilon_{ij} = u_{i,j}+u_{j,i}$. These results are consistent with the analytical prediction of homogenized octet-truss elastic constant with facets aligned with the principal directions \cite{Fleck-Deshpande}. The octet-truss architected material is known to be a cubic orthotropic material with the same Young's modulus along the three orthotropy axes \cite{Fleck-Deshpande}. 

Let us, finally, remark that the inelastic behavior of the unit cell has been a subject of constitutive modeling in the past. 
In Ref.~\cite{Fleck-Deshpande}, the authors worked out an anisotropic yield criterion, whose limitations were openly admitted. 
In these circumstances, d-refinement can provide an efficient pathway to macroscale simulations while a fully-satisfactory constitutive model is being devised. 



\begin{figure}[H]
\centering
\captionsetup[subfigure]{justification=centering}
\begin{subfigure}[b]{.64\linewidth}
\includegraphics[width=\linewidth]{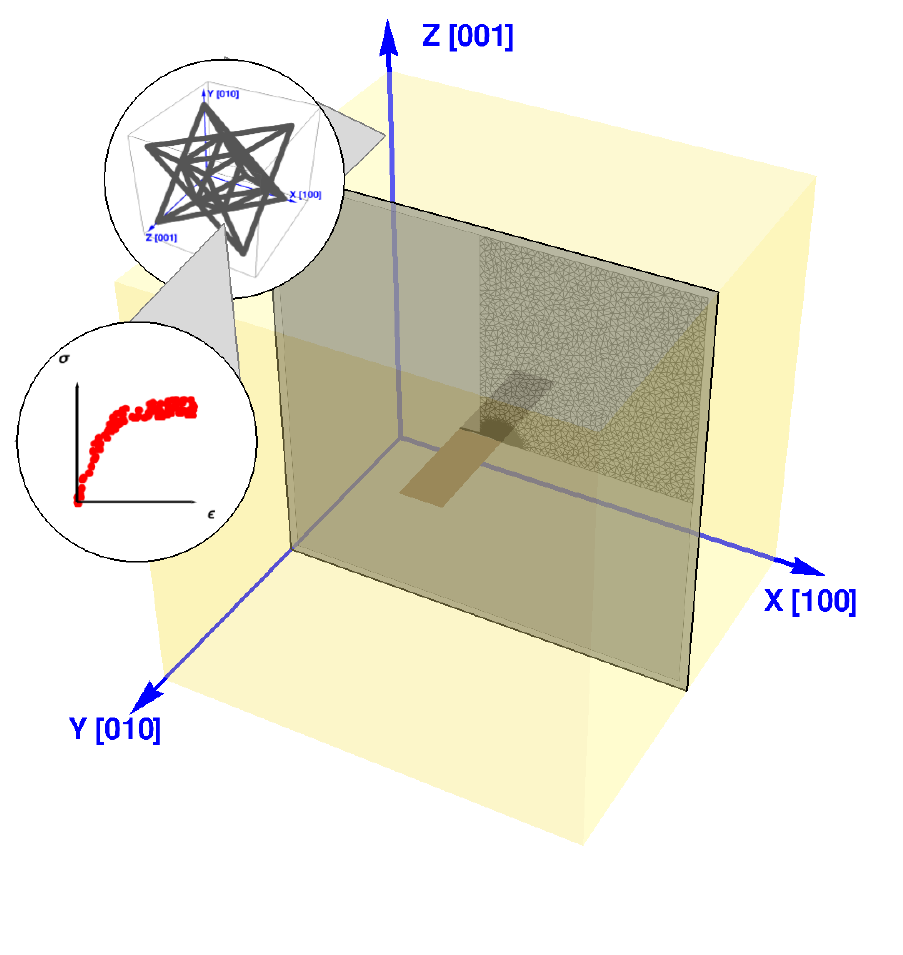}
\caption{\Large (a)}\label{fig:11a}
\end{subfigure}
\begin{subfigure}[b]{.35\linewidth}
\includegraphics[width=\linewidth]{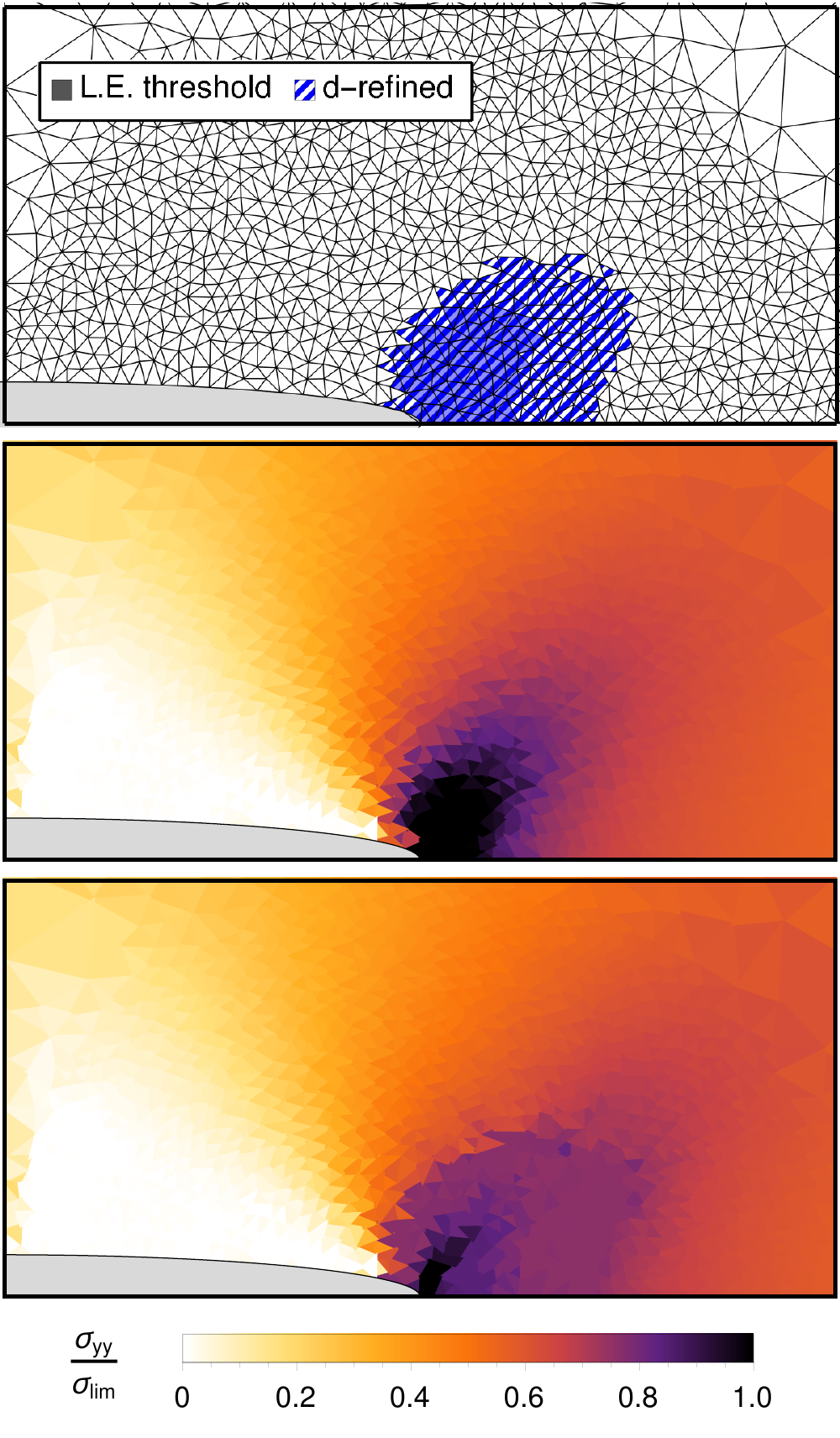}
\caption{\Large (b)}\label{fig:11b}
\end{subfigure}
    \caption{Macroscale simulation using d-refinement with data generated using RVE probing with DDCM. %
    (a) Scheme of cubic block of architected metamaterial containing a long crack within (cf. Ref.~\cite{toughness_metameterials}). The middle plane includes the reduced mesh used to generate plane-strain results. It also represents TMPTA dataset used to characterize the unit cell and a scheme of the octet-truss.
    (b) Results. Top: detail of the mesh around the crack tip including area above threshold according to linear-elastic pre-analysis (solid gray) and elements that became DD during the d-refinement solution (blue, diagonal hatching).
    Middle: dimensionless vertical normal stress $\sigma_{\text{yy}}$ (normalized by $ \sigma_{\text{max}} \approx 2p$) obtained with linear-elastic simulation (darkest zone is close to or above the critical value). 
    Bottom: dimensionless vertical normal stress obtained with d-refinement simulation. See how the stress concentration at the tip region is blurred.
    }
    \label{fig:test}
\end{figure}

\paragraph{Macroscale} The goal of this model is to study how the stress around a crack tip is distributed for an architected metamaterial. 
Linear-elastic fracture mechanics (LEFM) yields infinite stress at the crack tip \cite{Griffith}. In reality, there is a region of damaged material around the tip, the ``fracture process zone'' \cite{FPZ}, that prevents the appearance of this singularity by redistributing stress over a greater area. The relative size thereof defines the range of validity of LEFM: if this region characteristic length is of the order of the geometrical features of the system \cite{Hutchinson}, then a ductile fracture viewpoint should be adopted.
Assuming a long crack and uniform loading, only a cross-section is considered by taking the plane strain assumption, as pictured in \Cref{fig:test}. 
The unit cell is a cube of side $l = 750\, \mathrm{\mu m}$. 
The model is qualitatively similar to the one presented in  \Cref{Sec:validation_plate} (\Cref{fig:plate_mesh_results_a}) but with the shape of the hole changed to an elongated ellipse in order to better represent a thin crack tip. Specifically, the short radius is $0.1a$, $2a = 40 l$ being the crack length. The total area of the plate is $2 H \times 2 H $, with $H = 10 a$.

The mesh is refined around the crack tip, up to an element size of the order of the octet-truss unit cell. The problem was solved with the d-refinement method and each element was equipped with a dataset sifted according to the criterion $\sigma_{\text{yy}} > \sigma_{lim}=280 \, \mathrm{kPa}$, as this value ensures that the unit cell remains elastic (value obtained after inspecting uniaxial traction results of the unit cell, assuming it to be the most unfavourable case). The traction magnitude in this case is $p = 160 \, \mathrm{kPa}$. 

\paragraph{Database Generation} In both verification cases, the dataset was generated synthetically, assuming knowledge of the underlying constitutive law. For the rod element, the phase-space is only two-dimensional and thus it can easily be populated using only the estimated minimum and maximum strain sustained by a bar as bounds. 
For the 2D elements, the phase space is six-dimensional, meaning that to populate it with a similar density as the two-dimensional phase space, an exponent of three must be applied to the number of datapoints (i.e. a $10^9$ datapoints in a unit hypercube in 6D have the same density as 1000 datapoints in a unit 2D square). 
As this is hardly tractable, some knowledge of the expected strain paths must be obtained to constrain the region of interest in phase space. 
Previously, this information came from the resolution of the problem with non-linear FEM, thus giving the exact required strain paths. For the architected metamaterial, no underlying constitutive law is available. The chosen approach was to first solve the problem with a linear elastic FEM solver using the cubic material linear-elastic constants. 
%
%
%
The strain state of each element was recorded. Microscale RVE simulations were then run on 30 strain states linearly interpolated between the zero strain and 1.5 times the recorded strain states. Given that the simulation is composed of 4940 elements, this resulted in $4940 \times 29 + 1 = 143261$ points (the zero-strain states were not reproduced for each element), but only 173 surpasses the aforementioned threshold.

 \subsection{Results}
 
 The results are summarized in \Cref{fig:11b}. 
 The top image depicts the mesh around the crack, those elements that the linear-elastic solver predicted to go above the threshold appear in solid color. The refined elements are superimposed with hatched diagonal filling. As expected, the refinement subsumes the above-threshold elements; the extent of the former can be thought as an estimation of the fracture process in the architected metamaterial.

 The first solve using linear FEM yields the results on the middle plot, with a sharp stress concentration at the crack tip. 
 Then, on the bottom plot, the results with the multiscale d-refinement approach show a more diffuse stress field, with a zone of near-max stress ($\sigma_{\text{max}}$) around the crack tip.
 
\begin{figure}
    \centering
    \includegraphics[width=0.75\textwidth]{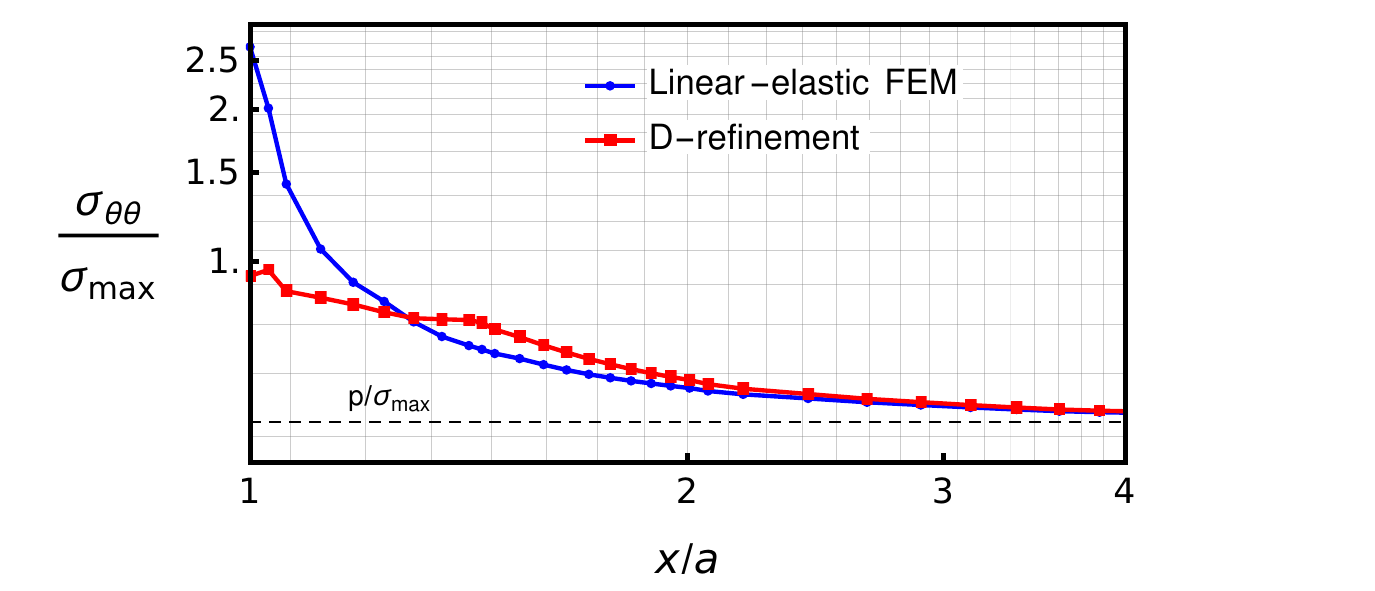}
    \caption{Hoop stress along the $x$-axis, from the crack tip ($x=a$) to the middle of the mesh (total horizontal extent of the plate from crack tip is $9a$).}
    \label{fig:stresses_crack}
\end{figure}

 The re-distribution of stress due to material non-linearity is also evident in \Cref{fig:stresses_crack}, where the hoop stress along the horizontal symmetry plane is visualized. 
 Both solutions converge to the remote loading conditions, but, as the crack is approached, the stress is raised faster in the multiscale d-refinement model than in the linear solution, until the latter overtakes the former around the tip. 
 The logic behind this behavior is well understood: the non-linear material cannot carry as much load around the tip, so more material is engaged in high-stress away from it. The stress does not plateau as in the elasto-plastic case, as the beyond-elastic stress states in the octect-truss can be diverse (\Cref{fig:strainstressRVE}). 

D-refinement could also allow to speedily assess fracture process zone sensitivity in terms of underlying unit-cell material. In recent experiments, the fracture initiation process was characterized for a truss made of similar yet brittle rods \cite{toughness_metameterials}. Considering this brittle micro-constituent behavior is straightforward in d-refinement, as the infinitesimal-strain moduli would not change and the non-linear behavior could be accounted for by erasing some rods from the unit cell, what in turn would translate in a different homogenized non-linear behavior.

\section{Conclusion}

The d-refinement method brings together the best of two computational paradigms: the speed and ease of definition and implementation of linear FEM method along with the ability of DDCM to faithfully capture complex material behavior. 
By resorting to DDCM only where and when necessary, efficient data-driven simulations can be run at little accuracy cost – when the material at hand displays a well-defined linear elastic regime at small-strain. 
The computational cost compares favorably to traditional incremental solvers such as Newton-Raphson without the need to define non-linear constitutive relations. 

For multiscale simulations, the offline database generation and data reuse are clear advantages against $FE^2$ methods. The application to architected meta-materials illustrates the potential use of efficient data-driven methods in multiscale scenarios. Now, thanks to DDCM and d-refinement, efficient simulations of systems featuring architected metamaterials are not contingent on prior development of a constitutive law informed by micromechanical behavior.

While the current implementation is restricted to path-independent (elastic) behavior in statics, the application of the d-refinement method could be readily extended to the formulation of DDCM in dynamics, large deformation or path-dependent material behavior. 


\section*{Acknowledgements}

S.W., J.-F. M. and J. G.-S. gratefully acknowledge the support of the Swiss National Science Foundation via grant "Wear across scales" (200021\_197152).

\section*{Supplementary Material}

Jupyter and Mathematica notebooks that implement the calculations described in the text are available from the last author GitHub page \texttt{github.com/jgarciasuarez}, in the repository named ``\texttt{d-refinement}''.

\newpage

\bibliography{bibliography}{}
\bibliographystyle{unsrt}



\end{document}